%% file: ChaI.tex
\documentclass[longauth]{aa} 

\usepackage{graphicx}
\usepackage{lscape}
\usepackage{longtable}
\usepackage{txfonts}
\begin{document}

   \title{The Gaia-ESO Survey: Structural and dynamical properties of the young cluster Chamaeleon I
   \thanks{This work is one of the last ones carried out with the help and support of our friend and 
   colleague Francesco Palla, who passed away on the 26/01/2016}
   \thanks{Based on observations 
   made with the ESO/VLT, at Paranal Observatory, under program 188.B-3002 
   (The Gaia-ESO Public Spectroscopic Survey)}}

   \author{G.~G.~Sacco \inst{\ref{inst1}} \and  L.~Spina\inst{\ref{inst1},\ref{inst2}} \and  S.~Randich\inst{\ref{inst1}} 
   \and  F.~Palla\inst{\ref{inst1}} 
   \and R.~J.~Parker \inst{\ref{inst3}\label{inst3a}} \and R.~D.~Jeffries \inst{\ref{inst4}} \and R.~Jackson \inst{\ref{inst4}} \and M.~R.~Meyer\inst{\ref{inst5}} 
   \and M.~Mapelli\inst{\ref{inst6}} \and A.~C.~Lanzafame\inst{\ref{inst8},\ref{inst7}} \and R.~Bonito\inst{\ref{inst9}} \and F.~Damiani\inst{\ref{inst9}} 
  \and E.~Franciosini \inst{\ref{inst1}} \and A.~Frasca\inst{\ref{inst7}} \and A.~Klutsch\inst{\ref{inst7}} \and L.~Prisinzano\inst{\ref{inst9}} 
   \and E.~Tognelli\inst{\ref{inst9a}} \and S.~Degl'Innocenti\inst{\ref{inst9a}} \and P.~G.~Prada~Moroni\inst{\ref{inst9a}} \and E.~J.~Alfaro\inst{\ref{inst10}} 
   \and G.~Micela\inst{\ref{inst9}} \and T.~Prusti\inst{\ref{inst11}} \and D.~Barrado\inst{\ref{inst12}} 
 \and K.~Biazzo\inst{\ref{inst7}} \and H.~Bouy\inst{\ref{inst12}} \and L.~Bravi\inst{\ref{inst1},\ref{inst13}} \and J.~Lopez-Santiago\inst{\ref{inst14}}
  \and N.~J.~Wright\inst{\ref{inst4}} \and A.~Bayo\inst{\ref{inst14a}}\and G.~Gilmore\inst{\ref{inst15}} \and A.~Bragaglia\inst{\ref{inst16}} \and E.~Flaccomio\inst{\ref{inst9}}  
 \and  S.~E. Koposov\inst{\ref{inst15}} \and  E.~Pancino\inst{\ref{inst1},\ref{inst17}} \and A.~R. Casey\inst{\ref{inst15}} 
  \and M.~T.~Costado\inst{\ref{inst10}} \and P.~Donati\inst{\ref{inst16}, \ref{inst18}} \and A.~Hourihane\inst{\ref{inst15}} 
   \and P. Jofr\'e\inst{\ref{inst15},\ref{inst19}}\and C.~Lardo\inst{\ref{inst15a}} \and J.~Lewis\inst{\ref{inst15}} \and L.~Magrini\inst{\ref{inst1}} 
   \and L.~Monaco\inst{\ref{inst21}} \and L.~Morbidelli\inst{\ref{inst1}} \and S.~Sousa\inst{\ref{inst22}} \and C.~C.~Worley\inst{\ref{inst15}} \and S.~Zaggia\inst{\ref{inst6}}
   }

   \institute{INAF-Osservatorio Astrofisico di Arcetri, Largo E. Fermi, 5, 50125, Firenze, Italy\label{inst1}
   \and Universidade de São Paulo, IAG, Departamento de Astronomia, Rua do Mãtao 1226, São Paulo, 05509-900, SP, Brasil\label{inst2}
  \and Department of Physics and Astronomy, University of Sheffield, The Hicks Building, Hounsfield Road, Sheffield, S3 7RH, UK\label{inst3}
   \and Royal Society Dorothy Hodgkin Fellow\label{inst3a}
   \and Astrophysics Group, Research Institute for the Environment, Physical Sciences and Applied Mathematics, Keele University,
    Keele, Staffordshire ST5 5BG, United Kingdom\label{inst4}
    \and Institute of Astronomy, ETH Zurich, Wolfgang-Pauli-Strasse 27, 8093 Zurich, Switzerland\label{inst5}
    \and INAF - Osservatorio Astronomico di Padova, Vicolo dell'Osservatorio 5, I35122, Padova\label{inst6} 
    \and Dipartimento di Fisica e Astronomia, Sezione Astrofisica, Universit\'{a} di Catania, via S. Sofia 78, 95123, Catania, Italy\label{inst8}
    \and INAF - Osservatorio Astrofisico di Catania, via S. Sofia 78, 95123, Catania, Italy\label{inst7}
   \and INAF - Osservatorio Astronomico di Palermo, Piazza del Parlamento 1, 90134, Palermo, Italy\label{inst9}
   \and Dipartimento di Fisica Universita' di Pisa ed INFN Sezione di Pisa, Largo B. Pontecorvo 3, 56127, Pisa, Italy\label{inst9a}
   \and Instituto de Astrof\'{i}sica de Andaluc\'{i}a-CSIC, Apdo. 3004, 18080, Granada, Spain\label{inst10}
   \and ESA, ESTEC, Keplerlaan 1, Po Box 299 2200 AG Noordwijk, The Netherlands\label{inst11}
  \and Depto. de Astrofísica, Centro de Astrobiología (CSIC-INTA), ESAC campus, 28691, Villanueva de la Cañada, Madrid, Spain\label{inst12}
  \and Universit\'a degli Studi di Firenze, Dipartimento di Fisica e Astrofisica, Sezione di Astronomia, Largo E. Fermi, 2, 50125, Firenze, Italy\label{inst13}
  \and Dpto. de Astrofísica y Cencias de la Atmósfera, Universidad Complutense de Madrid, 28040, Madrid, Spain\label{inst14}
  \and Instituto de F\'isica y Astronomi\'ia, Universidad de Valparai\'iso, Chile\label{inst14a}
  \and Institute of Astronomy, University of Cambridge, Madingley Road, Cambridge CB3 0HA, United Kingdom\label{inst15}
  \and Astrophysics Research Institute, Liverpool John Moores University, 146 Brownlow Hill, Liverpool L3 5RF, UK\label{inst15a}
  \and INAF - Osservatorio Astronomico di Bologna, via Ranzani 1, 40127, Bologna, Italy\label{inst16}
  \and ASI Science Data Center, Via del Politecnico SNC, 00133 Roma, Italy\label{inst17}
  \and Universit\`a di Bologna, Dipartimento di Fisica e Astronomia, viale Berti Pichat 6/2, 40127 Bologna, Italy\label{inst18}
  \and N\'ucleo de Astronom\'ia, Facultad de Ingenier\'ia, Universidad Diego Portales,  Av. Ejercito 441, Santiago, Chile\label{inst19}
  \and Departamento de Ciencias Fisicas, Universidad Andres Bello, Republica 220, Santiago, Chile\label{inst21}
  \and Instituto de Astrof\'isica e Ci\^encias do Espa\c{c}o, Universidade do Porto, CAUP, Rua das Estrelas, 4150-762 Porto, Portugal\label{inst22}
                    }
   \date{}

 \abstract
  { Investigating the physical mechanisms driving the dynamical evolution of young star clusters is fundamental to
   our understanding of the star formation process and the properties of the Galactic field stars. 
   The young ($\sim2$ Myr) and partially embedded cluster Chamaeleon I is one of the closest laboratories to study the 
   early stages of star cluster dynamics in a low-density environment. The aim of this work is to study the structural 
   and kinematical properties of this cluster combining parameters from the high-resolution
   spectroscopic observations of the Gaia-ESO Survey with data from the literature.
   Our main result is the evidence of a large discrepancy between the velocity dispersion
   ($\rm \sigma_{stars}= 1.14 \pm 0.35~km~s^{-1}$) of the stellar population
   and the dispersion of the pre-stellar cores ($\rm \sim 0.3~km~s^{-1}$) 
   derived from submillimeter observations. The origin of this discrepancy, which
   has been observed in other young star clusters is not clear. It has been suggested that 
   it may be due to either the effect of the magnetic field on the protostars and 
   the filaments, or to 
   the dynamical evolution of stars driven by two-body interactions.  
   Furthermore, the analysis of the kinematic properties of the stellar population put in evidence
    a significant velocity shift ($\rm \sim 1~km~s^{-1}$) 
   between the two sub-clusters located around the North and South main clouds of the cluster. 
   This result further supports a scenario, where clusters form
   from the evolution of multiple substructures rather than from a monolithic collapse.

   Using three independent spectroscopic indicators (the gravity indicator $\gamma$, 
   the equivalent width of the Li line at 6708 \AA, and the H$\alpha$ 10\% width), 
   we performed a new membership selection. We found six new cluster members all located in the outer region 
   of the cluster, proving that Chamaeleon I is probably more extended than previously thought. 
   Starting from the positions and masses of the cluster members, we derived the level of substructure $Q$, the
   surface density $\Sigma$ and the level of mass segregation $\Lambda_{MSR}$ of the cluster. The comparison between 
   these structural properties and the results of N-body simulations suggests that the cluster 
   formed in a low density environment, in virial equilibrium or supervirial, and highly substructured.
}

   \keywords{Stars: kinematics and dynamics -- Stars: pre-main sequence -- open clusters and associations: individual: Chamaeleon I -- Techniques: spectroscopic}
\titlerunning{The Gaia-ESO Survey: Structure and dynamics of Chamaeleon I}

   \maketitle
%

\section{Introduction}

The majority of stars do not form in isolation, but in clusters 
following the fragmentation and collapse of giant molecular clouds
\citep{Lada:2003, Mckee:2007}. Studying 
the formation and evolution of young clusters 
is fundamental to understand the star formation process 
and the properties of stars and planetary systems observed in the  Galactic field,
since they may depend on the formation environment
\citep[e.g.,][]{Johnstone:1998, Parker:2009, Rosotti:2014}.

Despite the large number of multi-wavelength observations of nearby star forming regions
carried out during the last two decades \citep[e.g.,][]{Carpenter:2000, Getman:2005, Gudel:2007,
Gutermuth:2009, Feigelson:2013}, the scientific debate on the initial conditions 
(i.e., stellar density, level of substructure, level of mass segregation) of star clusters
and on the mechanisms driving the dissolution of most of them within 10 Myr is still open.
In particular, it is not clear if the majority of stars form in very dense 
($\rm \ga 10^4 stars~pc^{-3}$) and mass segregated clusters
\citep[e.g.,][]{Kroupa:2001, Banerjee:2014}, or in a hierarchically structured environment
spanning a large density range \citep[e.g.,][]{Elmegreen:2008, Bressert:2010}.
Furthermore, the cluster dispersion may be triggered by gas expulsion due to the feedback of high-mass stars
\citep[e.g.,][]{Goodwin:2006, Baumgardt:2007},
or the dynamical evolution of star clusters could be driven by two-body interactions 
and the effect of the feedback may not be relevant \citep[e.g.,][]{Parker:2013a, Wright:2014}.

From the observational point of view, the main requirements to solve this
debate are: a) an unbiased census of young stellar populations in 
star-forming regions spanning a large range of properties 
(i.e., density, age, total mass); b) the determination of the structural properties
of young clusters (density, level of substructure and mass segregation)
based on robust statistical methods that can be used for comparison with models;
c) precise measurements of stellar velocities that allow us to
resolve the internal dynamics of clusters and derive their
dynamical status (e.g., virial ratio, velocity gradients).

During the last few years, progress has been achieved thanks to the 
theoretical efforts dedicated to better define the structural properties of clusters,
\citep[e.g.,][]{Cartwright:2004, Allison:2009, Parker:2012a}, to the comparison
between models and observations \citep{Sanchez:2009, Parker:2011, Parker:2012b, Wright:2014, Da-Rio:2014, 
Mapelli:2015}, and to dedicated observational studies of the dynamical properties of young clusters 
based on accurate radial velocities (RVs) \citep{Furesz:2006, Furesz:2008, Cottaar:2012a, 
Jeffries:2014, Foster:2015, Sacco:2015, Tobin:2015, Rigliaco:2016, Stutz:2016}. However, it is essential to extend these studies to a
large number of clusters to cover the full space of relevant physical parameters (e.g., number of stars, stellar
density, and age).

The young cluster Chamaeleon I (Cha I) is located around one of the 
dark clouds of the Chamaeleon star forming complex (see \citealt{Luhman:2008a}
for an exhaustive review). It is the ideal laboratory to study 
the formation and early evolution of a low-mass cluster due to its 
proximity (distance=$160\pm15$ pc; \citealt{Whittet:1997}), the presence of a molecular cloud actively forming stars, and
a stellar population composed of $\sim$240 members \citep{Luhman:2008a, Tsitali:2015} distributed over 
an area of a few square parsecs.

The Cha I molecular cloud has been studied by \cite{Cambresy:1997}, who obtained an extinction map 
up to $A_V\sim10$ mag, and by radio surveys of C$^{18}$O or $^{12}$CO emission
(e.g., \citealt{Boulanger:1998, Mizuno:1999, Mizuno:2001, Haikala:2005}).
In particular, \cite{Mizuno:2001} used the $^{12}$CO to estimate the total mass of the 
cloud ($\sim1000~\rm M_{\sun}$), while \cite{Haikala:2005} mapped the structure of the 
filaments by observing the C$^{18}$O emission. The filaments follow the structure of the cloud that
is elongated in the NW-SE direction and perpendicular to the magnetic field.
Protostellar cores within the cloud have been identified by \cite{Belloche:2011} observing 
the continuum emission at 870 $\mu m$ and, more recently, studied by \cite{Tsitali:2015}
in several molecular transitions. \cite{Tsitali:2015} measured the velocity dispersion of the cores ($\sim0.3 \rm ~km~s^{-1}$)
and conclude that their dynamical evolution is not affected by 
interaction and competitive accretion, since the collisional timescale is much 
longer than the core lifetime.

Several multi-wavelength studies have been carried out to identify the 
stellar and brown dwarf population of Cha I (e.g. \citealt{Carpenter:2002, Luhman:2004, Luhman:2008, Luhman:2008a, 
Lopez-Marti:2013, Lopez-Marti:2013a}, and references therein). \cite{Luhman:2008a} compiled a list of 237 members using
many membership indicators such as: the position in the HR diagram, high optical extinction, 
intermediate gravity between giants and main sequence stars, the presence of the 
Li absorption line at 6708 \AA, infrared excess emission, the presence of emission lines,
proper motions and RVs. From the position in the HR diagram, \cite{Luhman:2007}
derived a median age of 2 Myr and suggested that Cha I is divided in two sub-clusters, one concentrated 
in the northern part of the cloud ($\delta > -77^{\circ}$) and one in the south ($\delta < -77^{\circ}$).
The former started to form stars 6 Myr ago, while the latter started later (4 Myr ago) and retains a larger amount of 
gas mass. Furthermore, they calculated an upper limit to the star formation efficiency of $\sim$10\%.
Several studies have been dedicated to measure the RVs and study the kinematics 
of the stellar and substellar populations of Cha I (\citealt{Dubath:1996, Covino:1997, 
Joergens:2001, Joergens:2006, Guenther:2007}). In particular, \cite{Joergens:2006} measured 
a RV dispersion of 1.2 $\rm km~s^{-1}$, but this result is based on 25 stars only.

Cha I has been one of the first young clusters observed by the Gaia-ESO Survey (GES). GES is a large
public spectroscopic survey carried out with the multi-object instrument FLAMES at the VLT, which feeds the
medium- and high-resolution spectrographs GIRAFFE and UVES. The main
goal of the survey is to derive RVs, stellar parameters (i.e., effective temperature, gravity, 
metallicity) and chemical abundances of 10$^5$ Milky Way stars in the 
field and in clusters \citep{Gilmore:2012, Randich:2013}. 
A study of stellar activity, rotation and accretion based on the GES observations of Cha I is reported in \cite{Frasca:2015}, 
while the iron abundances of a selected sample of stars observed at high resolution are
reported in \cite{Spina:2014}. Here, we investigate the structural and dynamical properties of Cha I
combining the new GES results with data available in the literature. The paper is organized as follows: 
in Sect. 2, we present the method used for selecting the targets, the observations, and the data retrieved from the GES
archive; in Sect. 3 we describe how we select the cluster members; in Sect. 4, we derive its structural properties; 
in Sect. 5, we derive the dynamical properties of the cluster;  in Sect. 6, we discuss our results on the
basis of the current models describing the dynamical evolution of low-mass star clusters, and in Section 7, we summarize the results
and draw our conclusions.

\input{Tab1.tex}

\begin{figure}
   \centering
   \includegraphics[width=9.5 cm]{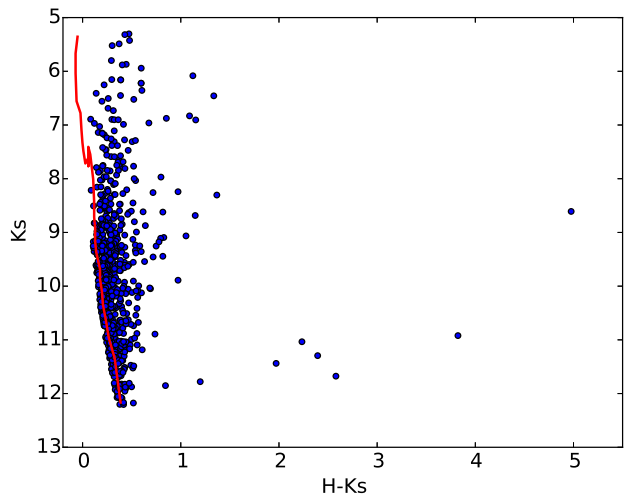}
      \caption{Color magnitude diagram of the observed targets in the Cha I region based on photometry
      from the Two Micron All Sky Survey \citep{Skrutskie:2006}, with overplotted the 10 Myr isochrone from the
      \cite{Siess:2000} models.}
         \label{fig:cmd_target}
\end{figure}


\section{Target selection, observations and data}

The target selection and the fiber allocation procedure have been carried out independently
for each cluster observed during the survey, however in order to maintain the
homogeneity within the GES dataset we followed common guidelines described in Bragaglia et al. (2017, in preparation).

\begin{figure}
   \centering
   \includegraphics[width=9.0 cm]{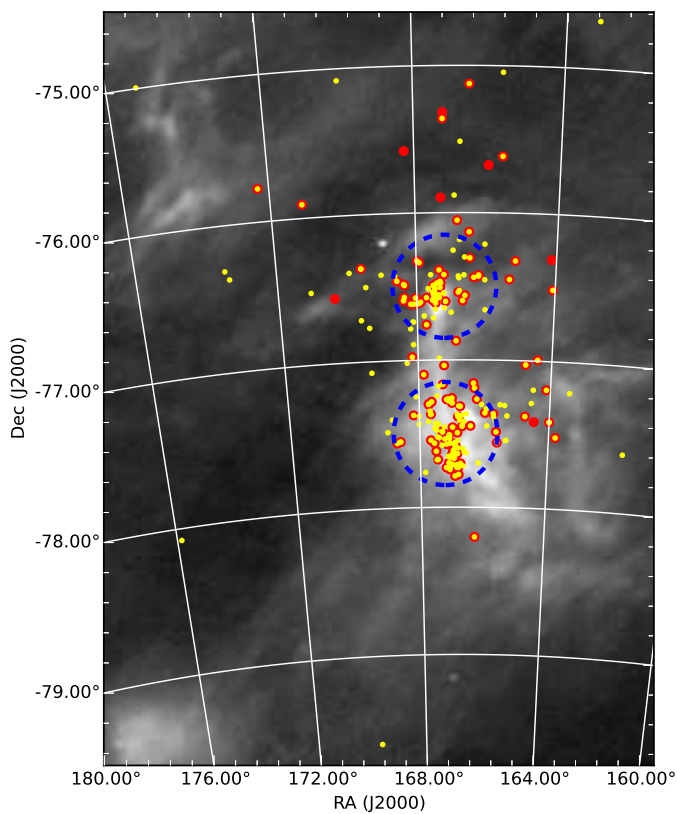}
      \caption{Far infrared (140 $\mu m$) map of the region around the young cluster Cha I 
      from the {\it AKARI} all-sky survey \citep{Doi:2015}.
      Yellow dots indicate the positions of all the known members from
      the literature, while the red bigger dots indicate the positions of all the members selected 
      by the GES observations according to the criteria discussed in Sect. \ref{sect:memb}.
      The dashed blue circles (centers Ra1 = 167.2$^{\circ}$ , Dec1= -76.5$^{\circ}$, 
      Ra2 = 167.2$^{\circ}$, Dec2 = -77.5$^{\circ}$ and radius 0.35 $^{\circ}$)
       delimit the north and south sub-clusters 
      (see Sect. \ref{Sect:substruct} and Sect. \ref{sect:subclusters}).}
         \label{fig:map}
\end{figure}

The selection of the targets for the observations of Cha I have been mostly based on the
infrared photometry from the 2 micron all sky survey (2MASS, \citealt{Skrutskie:2006}), since
optical photometric catalogues available in the literature are incomplete and not
homogeneous. The target selection and the fiber allocation process can be divided in two steps: 
we first compiled a list of candidate members in the region of the sky around the cloud
(10:45~$\leq$~RA~$\leq$~11:30 and -79:00~$\leq$~DEC~$\leq$~-75:00), then we 
defined the position of the FLAMES fields of view (FOVs, diameter 25$\arcmin$) and we allocated the 
largest possible number of fibers on candiate members.

For compiling the list of candidate members, we collected all the 2MASS sources  
with an optical counterpart from the Tycho 2 or USNO-B1 \citep{Hog:2000,Monet:2003} catalogues brighter than R=17, which
corresponds to the magnitude limit of the survey (V=19) for very low-mass stars.
Then, from this list we selected only the sources that in a K vs. H-K color-magnitude 
diagram are located above the 10 Myr isochrone retrieved from the \cite{Siess:2000} evolutionary models 
(see Fig. \ref{fig:cmd_target}). 
Using this method, we compiled a list of 
1933 candidate members.
On the basis of this list and the positions of the known members, we chose 25 
FOVs. Many of them were located along the main cloud, where most
of the known members of the cluster are distributed, while a few FOVs were located on 
the outer regions, with the aim of looking for new members in regions which are
 poorly studied (the structure of the cloud and the positions of the known members are shown in Fig. \ref{fig:map}). 
FOVs in the outer regions have been chosen in order to cover each latitude
and longitude around the main cloud, focusing on the regions with higher spatial density of 
sources. 

We observed a total of 674 stars with GIRAFFE and 49 with UVES
(3 in common between the two spectrographs), of which 113 are known members of the cluster
on the basis of catalogues reported in the literature \citep{Luhman:2004a, Luhman:2007, Luhman:2008}. 
Most of the known members have been excluded because too faint to be observed with FLAMES. 
Several fibres have been allocated to the sky in order to allow a good background subtraction.
Observations have been carried out during three different runs between
March and May 2012\footnote{Technical details on the fibre allocation procedure and 
observations are discussed in Bragaglia et al. (2017, in preparation)}, 
using the HR15N setup (R$\sim$17,000, $\Delta \lambda$=647-679 nm) for GIRAFFE and the 
580 nm (R$\sim$47,000, $\Delta \lambda$=480-680 nm) setup for UVES. The median signal-to-noise
ratio of the final spectra is 58 and 62 for GIRAFFE and UVES spectra, respectively.

All GES data are reduced and analysed using common methodologies and
software to produce an uniform set of spectra and stellar parameters, which is periodically
released to all the members of the consortium via a science archive\footnote{The GES science archive is run by the Royal Observatory
of Edinburgh. More information on the archive are available at the website ges.roe.ac.uk}. 
In this paper, we use only data from the third internal data releases (GESviDR3) of February 2015,
with the exception of errors on the GIRAFFE RVs, which are calculated on the basis of the
empirical formulae provided by \cite{Jackson:2015}.

The methodologies used for the data reduction and the derivation of RVs are described in section 
2.2 and 2.3 of \cite{Jeffries:2014} for the GIRAFFE data, and in \cite{Sacco:2014a} for the UVES data.
\cite{Lanzafame:2015} describes in details the procedures used to derive
the stellar paramaters (i.e., effective temperature and gravities), accretion indicators (e.g.,
H$\alpha$ width at 10\% of the peak -- H$\alpha$10\%), and the equivalent width of the Li line at 6708 
\AA~(hereafter EW(Li)).

\section{Membership selection \label{sect:memb}}

A detailed selection of members among the stars observed with UVES 
has been carried out by \cite{Spina:2014}, who confirmed all the known members from the literature and
did not find any new members, therefore we will focus only on stars observed with GIRAFFE.

\begin{figure}
\centering
\includegraphics[width=8.0 cm]{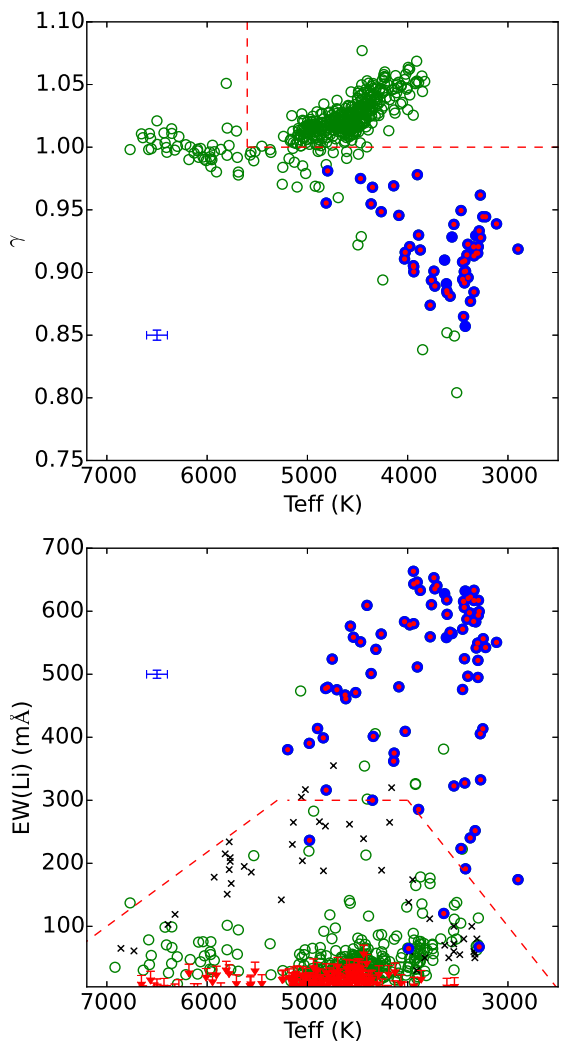}
\caption{The upper and the lower panels show the gravity index $\gamma$ and EW(Li) 
as a function of the stellar effective temperature, respectively. 
The index $\gamma$ and the EW(Li) measured from the GES spectra are indicated with circles:
the empty green circles are the non-members, the filled blue circles are the cluster members, and the red dots represent
the members already known from previous studies. Upper limits are reported with red downwards arrows. 
Median error bars are reported on the left for both panels.
The black crosses in the bottom panel are the EW(Li) measured for stars in the 30-50 Myr open cluster IC 2602 by 
\cite{Randich:1997, Randich:2001}. In both panels, the dashed red lines indicate the threshold used to separate members 
and non-members. In the bottom panel, a few stars are reported as non-members despite being above the dashed red line because 
they have been excluded according to the criterium based on the gravity index, and other stars are classified as members despite
being below the dashed line, because they are strong accretors, as discussed in Sect. \ref{sect:memb}.}
\label{fig:memb_plots}
\end{figure}

Since all the stars formed in the same region have very similar velocities, spectroscopic measurements 
of the RVs are often considered one of the most robust tools to select the members of a cluster. 
However, one the main goals of this work is to study the dynamical properties of Cha I
(e.g, the RV dispersion, the presence of multiple populations) therefore,
we will use the RVs only to discard the obvious non-members, namely the stars outside of the range
$0< RV < 30~\rm km~s^{-1}$, and for a more accurate selection of the cluster members, we will 
use three independent spectroscopic parameters included in the GES database: the gravity index $\gamma$, the EW(Li) and the H$\alpha$10\%.
The major source of contamination within a sample of candidate members of a nearby young cluster selected on
the basis of photometric data are background giants. The giants can be identified using the surface gravity
index $\gamma$, defined by \cite{Damiani:2014} with the specific goal of measuring gravities using the 
GES GIRAFFE spectra observed with the HR15N setup. The upper panel of Fig. \ref{fig:memb_plots} shows $\gamma$ 
as a function of the effective temperature for the GES targets observed in Cha I.
The locus of the giant stars is clearly visible in the upper part of the plot and well separated from 
the main sequence and pre-main sequence stars. Similarly to previous works 
\citep{Prisinzano:2016, Damiani:2014}, we classified as giants all stars with an effective temperature
lower than 5600 K and $\gamma > 1$.

After the giants stars have been excluded, we need to exclude stars
older than Cha I located in the foreground. The most powerful tool to perform this selection is the EW(Li), 
since late-type stars rapidly deplete their photospheric lithium after 5-30 Myr \citep[e.g.,][]{Soderblom:2010}.
At constant age, the EW(Li) depends on the effective temperature, therefore 
we cannot define a single threshold for the whole sample. 
The lower panel of Fig. \ref{fig:memb_plots} shows the EW(Li) as a function of $T_{eff}$
for the observed stars in Cha I and for the stars of the 30-50 Myr open cluster IC 2602 observed by 
\cite{Randich:1997, Randich:2001}. We select as cluster members all the stars with EW(Li)
above the dashed line in the lower panel of Fig. \ref{fig:memb_plots}, which represents the upper envelope 
of the of the EW(Li) measured for the stars belonging to IC 2602. In a few cases, when the EW(Li) but not the 
effective temperature have been derived from the GES spectra (see \citealt{Lanzafame:2015} for details), we assume
the highest threshold (EW(Li)=300 \AA).
However, the EW(Li) can be underestimated in stars with a very strong mass accretion rate due to 
the continuum emission in excess with respect to
the photospheric one produced by the accretion shock \citep[e.g.,][]{Palla:2005, Sacco:2007}, therefore, 
we include in the sample of members all the stars that can be classified as accretors according to the 
criterium based on the width at 10\% of the peak of the $H\alpha$ line
($H\alpha10\%>270\rm~km~s^{-1}$) defined by \cite{White:2003}. In the bottom panel of Fig. \ref{fig:memb_plots}
the cluster members are indicated with filled blue circles. The members below the dashed line have been included because 
of the $H\alpha10\%$ while the non-members above the line have been excluded because of the gravity index $\gamma$.

Using these criteria, we selected as members of Cha I 89 stars observed with GIRAFFE. This sample includes
7 new members and 82 known members from the literature. Fourteen known members do not meet the membership criteria.
Seven of them have been excluded because they are out of the range of RVs ($0< RV< 30~\rm km~s^{-1}$). However,
all these stars are strong accretors (H$\alpha10\% >300 ~\rm km~s^{-1}$), therefore the RV derived by the GES pipeline
could be wrong due to the presence of strong emission lines produced by material moving at a different velocity with respect
to the photosphere. We will use these stars for the analysis of the structural properties of the cluster 
(see Sect. \ref{sect:struct}), but we will exclude them 
from the analysis of the dynamical properties, which is based on the RVs.
Six known members have been excluded because the SNR$<10$ is too low to derive EW(Li) and H$\alpha$10\% 
\citep[see][]{Lanzafame:2015}, but there is no evidence suggesting that these stars are not members,
therefore, we will include them in the final catalogue. Finally, the star HD 97300 is too hot (SpT= B9) to exhibit 
the Li absorption feature at 6708 \AA, but it is surrounded by a ring of dust due to a bubble blown by the star 
\citep{Kospal:2012}. 
This proves that it belongs to the Cha I star forming region and can be included in the catalogue of members used
for the analysis discussed in the following sections.
To summarize, the final catalogue of members includes 103 stars (96 already known and 7 new) observed with GIRAFFE 
and 17 stars observed with UVES (discussed in \citealt{Spina:2014}) for a total of 120 members.
We note that our analysis proves that the 7 new members are young stars, but does not demonstrate that
belong to Cha I, since they could be members of the two young stellar associations $\epsilon$ Cha and 
$\eta$ Cha, which are slightly older (4-9 Myr, \citealt{Torres:2008}), have a similar radial velocities 
and are located in the foreground of Cha I. As shown by \cite{Lopez-Marti:2013a} 
(see Fig. 1 of that paper), the two associations can be easily separated from Cha I using tangential 
motions, so we retrieved the proper motions of these stars from the UCAC 4 \citep{Zacharias:2013} and, when 
not available, from the PPMXL \citep{Roeser:2010} catalogues. We found that all the 
new members have tangential motions consistent with the Cha I cluster except one star (GES ID. 10563146-7618334), 
which is closer to the $\epsilon$ Cha association.

The list of cluster members and the data used for the membership 
selection are reported in Table \ref{tab:members}, while their positions are plotted on the map in Fig. \ref{fig:map}
(red dots) together with all the known members of the clusters compiled from the literature (yellow dots).
The number of new members does not significantly increase the population of Cha I.
However, they belong to the sparse population located in the outer region surrounding the main cloud. 
So our study proves that this outer population is richer than previously thought.

\section{Structural properties \label{sect:struct}}

As proved by several studies \citep[e.g.,][]{Scally:2002, Schmeja:2008, Allison:2010, Moeckel:2010, Malmberg:2011, 
Kruijssen:2012, Parker:2014}, a knowledge of the structural properties of open clusters is
fundamental to understand their origin, their dynamical evolution and the effects of the star formation environment
on the properties of stars and planetary systems. In this work we will focus on three structural properties:
the level of substructure, the stellar density and the mass segregation. 

\subsection{Sample and stellar masses}

The sample of stars used for the structural analysis includes all the previously known members 
(observed or not by GES) and the new members discovered by GES. 
We exclude stars with $A_{J}>1.2$, because catalogues available in the literature are not complete 
for higher extinction \citep{Luhman:2007}.

\begin{figure}[hbt]
\centering
\includegraphics[width=8.0 cm]{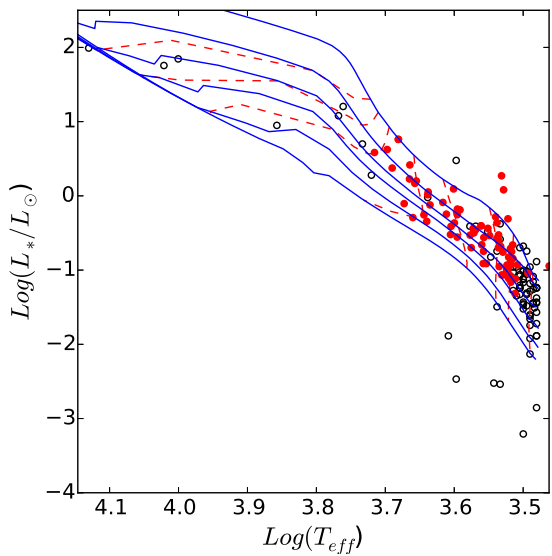}
\caption{HR diagram of members in Cha I selected from the literature and the GES data with $A_{J} < 1.2$ and 
$T_{eff} > 3000~\rm K$. Temperatures and luminosities have been derived from the GES spectra and 
the 2MASS photometry for the red dots, and from the literature for the other stars. Isochrones (at 0.5, 1.5, 3.0, 5.0, 10.0, and 20.0 Myr) and tracks (at 0.1, 0.2, 0.3, 0.6, 1.0, 
1.8, 2.4, 3.0 solar masses) from an improved version of the \cite{Tognelli:2011, Tognelli:2012} pre-main sequence 
evolutionary models are reported with continuous blue and dashed red lines, respectively.
}
\label{fig:HR_diag}
\end{figure}

For the analysis of the mass segregation, we derived homogeneous estimates of stellar masses 
from the positions of stars in the HR diagram plotted in Fig. \ref{fig:HR_diag},
using the pre-main sequence evolutionary models developed by \cite{Tognelli:2011,Tognelli:2012}.
We decide to use this set of models instead of that provided by \cite{Siess:2000} and recommended by 
the Gaia-ESO guidelines for the target selection process, because they have been more recently updated.
No masses have been estimated for stars cooler than 3000 K, younger than 0.5 Myr and older than 20 Myr 
(i.e., above/below the upper/lower isochrones plotted in the HR diagram). Very cool and very young stars 
have been excluded because mass estimation based on pre-main sequence evolutionary models could be very uncertain.
Stars located below the 20 Myr isochrone have been excluded because, as suggested by \cite{Luhman:2007, Luhman:2008b},
their luminosity is underestimated due to the presence of a circumstellar disk seen edge-on, which absorbs most 
of the photospheric emission.

\begin{figure}[hbt]
\centering
\includegraphics[width=8.0 cm]{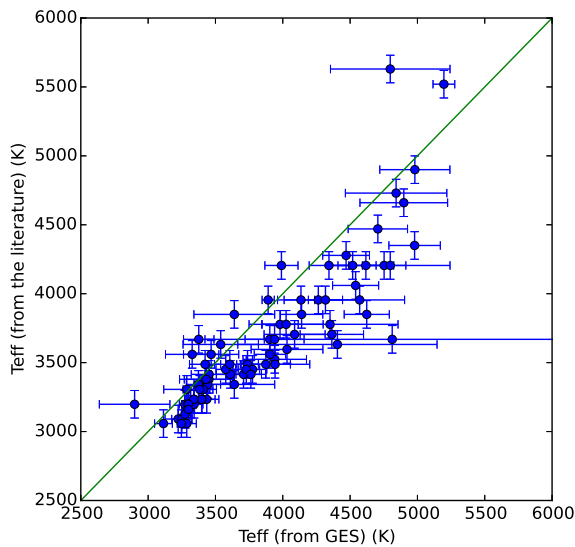}
\caption{Comparison between the effective temperature retrieved from the literature (y-axis) and 
measured from the GES spectra (x-axis).
}
\label{fig:Teff_comparison}
\end{figure}

To build the HR diagram we use 2MASS photometry and the GES parameters when available, otherwise  
we use the parameters from \cite{Luhman:2004a} and \cite{Luhman:2007}.
Specifically, the effective temperatures have been directly measured from the GIRAFFE and UVES spectra 
\citep{Lanzafame:2015} in the GES sample and from low resolution spectra for the other stars 
\citep{Luhman:2004a,Luhman:2007}. A comparison of the effective temperatures derived by the GES
spectra and those obtained from low resolution spectroscopy is reported in 
Fig. \ref{fig:Teff_comparison}, which shows that the latter are 
slightly lower in the range between 3600 and 4600 K. We believe that this 
systematic discrepancy can be due to either the scale used by \cite{Luhman:2004a,Luhman:2007}
to convert spectral types into temperature (GES temperature are derived directly from the spectra)
or to the different spectroscopic resolution of the spectra used for the analysis.

Luminosities have been derived from the 2MASS J magnitude corrected
for absorption, using the bolometric correction reported in \cite{Pecaut:2013} and assuming a distance 
of 160 pc. For the GES sample, we calculate the absorption $A_J$ from $E(J-H)=(J-H)-(J-H)_0$, assuming 
$A_J/E(J-H)=0.38$ as in \cite{Luhman:2004a}. The intrisic color $(J-H)_0$ was calculated from the effective 
temperature using the color-temperature trasformations from \cite{Pecaut:2013}. 
For stars that do not belong to GES catalogues, the infrared absorption has been retrieved 
from \cite{Luhman:2007, Luhman:2004}. The masses of the stars used for the structural analysis are reported in table 
\ref{tab:struct}.

To understand if the use of different catalogues for the estimation of masses can affect our results, 
we performed all our analysis using only data from the literature for all the known members finding no 
relevant differences. We stress that in the context of this paper the stellar masses are only used 
for studying the level of the mass segregation (see Sect. \ref{sect:stellar_density} and \ref{sect:mass_seg}). 
For this scope, we do not need the specific values of the masses but only to put them in order from the most 
to the less massive.

\input{Tab2.tex}

\begin{figure*}
   \centering
   \includegraphics[width=14 cm]{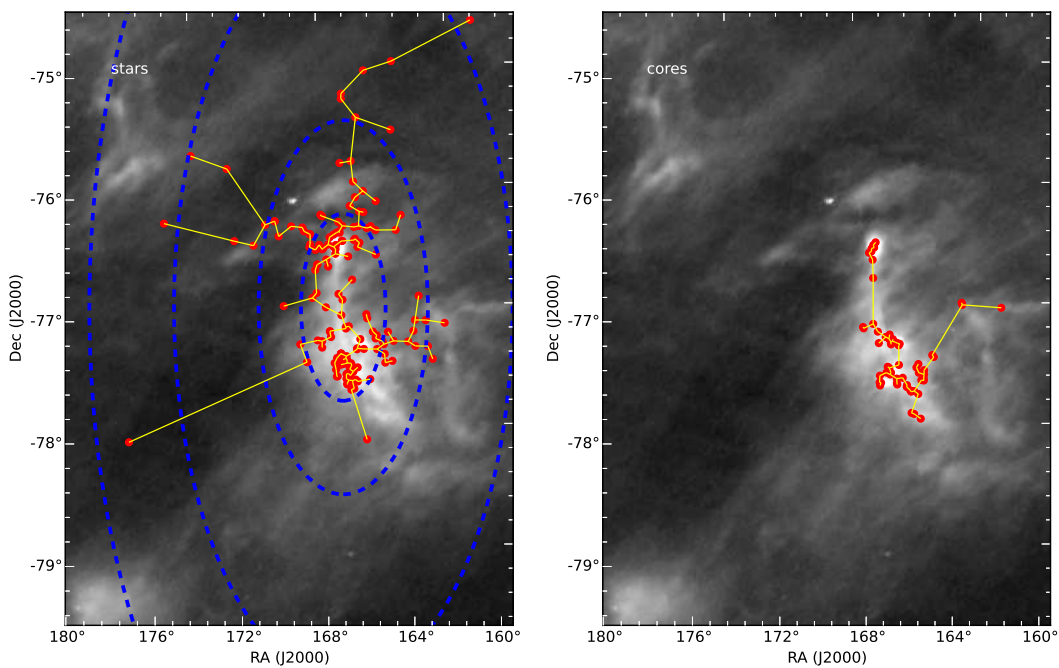}
      \caption{Spatial distribution of the stars (left panel) and starless cores (right panel)
      used to calculate the $Q$ parameter. The minimum spanning tree is plotted with a yellow line. 
      The blue dashed lines in the left panel show the elliptic boundaries of the regions including the stars used to 
      calculate the values of the $Q$ parameter reported in Fig. \ref{fig:Q_axis}.}
         \label{fig:mst}
\end{figure*}

\subsection{Level of substructure \label{Sect:substruct}}

To measure the level of substructure, we use the $Q$-parameter introduced by \cite{Cartwright:2004}. 
This parameter is defined as the ratio:

\begin{equation}
Q=\frac{\bar{m}}{\bar{s}}
\end{equation}

\noindent where $\bar{m}$ is the mean length of the edges of the minimum spanning tree 
(MST) connecting the stars, normalized by the factor $(NA)^{1/2}/(N-1)$ (N is the total 
number of stars and A is the area of the cluster), and $\bar{s}$ is the mean separation between 
the stars divided by the cluster radius\footnote{The radius and the 
area of the cluster are calculated as described in \cite{Cartwright:2004}. 
Specifically the former is the distance
beween the center of the cluster and the most distant stars, and the latter 
is the area of a circular surface 
with the same radius.}. Several simulations demonstrate that clusters with $Q > 0.8$ are characterized 
by a smooth radially concentrated structure,
which is probably the result of dynamical evolution occurring after the cluster formation,
while clusters with $Q < 0.8$ are characterized by a high level of substructure
and closely resemble the filamentary structure of the molecular clouds where they formed 
\citep[e.g., ][]{Schmeja:2006,Parker:2012a}.
To calculate the best estimator $\hat{Q}$ and the standard error $\sigma (\hat{Q})$ of the $Q-$parameter,
we use the \textit{Jackknife} method \citep{Quenouille:1949,Tukey:1958}. This is a resampling method that consists
in calculating the value of $Q_i$ for $N$ different samples composed of all the stars except for the $i-th$ star 
(with $i$ from 1 to N). The best estimator and the standard error are equal to:

\begin{equation}
\hat{Q}=\frac{1}{N} \sum \limits_{i=1}^N Q_i\\
\sigma (\hat{Q})=\sqrt{\frac{N-1}{N} \sum \limits_{i=1}^N (Q_i-\hat{Q})^2 }
\end{equation}

The relation between $Q$ and the level of substructure
has been calculated by \cite{Cartwright:2004} through simulations considering isotropic clusters. 
However, \cite{Cartwright:2009} pointed out that in elongated clusters 
$Q$ could be biased towards lower values and estimated a correction
factor which depends on the cluster aspect ratio. 
Since Cha I is characterized by a slightly elongated structure, we 
applied these corrections on our results.

For our sample, the resulting $\hat Q = 0.80\pm 0.08$ is higher than what found by previous studies 
carried out by \cite{Cartwright:2004} and \cite{Schmeja:2006} ($Q$ = 0.68 and 0.67, respectively). 
\cite{Cartwright:2004} did not consider the elongation in their calculation, but this do
not explain such a large discrepancy, since the elongation factor for our sample is just 1.03.
Therefore, this discrepancy is most likely due to the different sample of members used for the calculations. 
In fact, \cite{Cartwright:2004} used the sample of members selected by \cite{Lawson:1996} and \cite{Ghez:1997}, 
while \cite{Schmeja:2006} results are based on a sample of members retrieved from \cite{Cambresy:1998}.
Both sample are less complete and cover a smaller area of the sky than ours. 
As discussed in the previous section, Cha I is composed of an inner denser region characterized by 
the presence of a molecular cloud still forming stars and an outer sparser region with no gas. 
Furthermore, our discovery of new members only in the outer part of the cluster proves that the level of completeness
of the star catalogue in the outer region is lower than in the inner one. 
To understand how this can affect the $Q$-parameter, we calculated 
$Q$ for four different samples composed of stars enclosed within the area of the sky 
delimited by ellipses with the same center ($RA = 167.2^{\circ}$, $DEC= -77.1^{\circ}$) and eccentricities 
$e= 0.89$, but different semi-major axes (see table \ref{tab:ellipses} and the left panel of Fig. \ref{fig:mst}).
In particular, the smallest ellipse contains only the stars in the inner embedded region, the largest 
includes all the stars and the two intermediate ones have a semi-major axes twice and four times the smallest.

\begin{figure}
   \centering
   \includegraphics[width=7.5 cm]{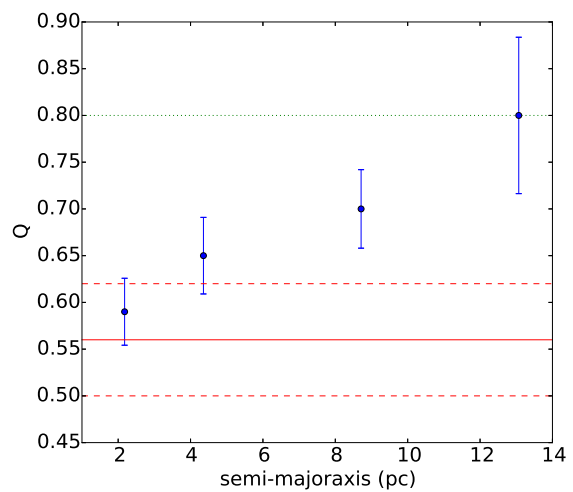}
      \caption{The $Q-$parameter as function of the semi-major axes of the ellipses, shown in Fig. \ref{fig:mst},
      which delimit the area enclosing all the stars used for the calculation. The dotted green line indicates the value of
      $Q$ expected for a sample of stars randomly distributed, while the continuous and the dashed red lines indicate the 
       the $Q-$parameter with error bars calculated from the positions of the pre-stellar cores.}
         \label{fig:Q_axis}
\end{figure}

\begin{table}
\caption{Properties of the ellipses used for investigating the relation between $Q$ and 
the completeness of the member sample\label{tab:ellipses}.}
\centering
\begin{tabular}{ccc}
\hline\hline
Semi-major axis &  N$_{star}$  & $Q$ \\
(degree)            &              &  \\
\hline
0.78 & 98  & 0.59$\pm$0.04 \\
1.56 & 143 & 0.65$\pm$0.04 \\
3.12 & 157 & 0.70$\pm$0.04 \\
4.68 & 160 & 0.80$\pm$0.08 \\
\hline
\end{tabular}
\end{table}

The value of $Q$ as a function of the semi-major axes of the ellipses is shown in Fig. \ref{fig:Q_axis} 
 and reported in table \ref{tab:ellipses}. 
The $Q-$parameter gradually increases from $Q\sim 0.6$ -- 
when we consider only the stars in the inner and denser region of the cluster -- to $Q\sim 0.8$ for 
the full sample.
This clear correlation between the $Q-$parameter and the area of the cluster considered to 
perform the calculation may be due to one or more of the following reasons. a) The stars in the inner region are 
younger and located very close to where they formed, so $Q$ is similar to what expected at the initial stage of the cluster 
formation \citep[e.g.,][]{Parker:2014}, when the distribution of stars resambles the distribution 
of gas in filaments, while the stars in the outer regions migrated from their formation site, so
their spatial distribution has been randomized;
b) Cha I is composed of multiple populations with different structural properties. In the different
subsets defined by the ellipses, these populations have different weigths, so the 
value of $Q$ changes according to which population is weighed the most; 
c) The presence of patchy extinction in the inner region produces substructures, which will not be
present if the full sample of stars would be visible.
d) The member selection in the outer regions is not complete. Missing some of the members, we
can miss some of the substructures, so $Q$ increases, when we include the outer part of the cluster.

\cite{Luhman:2007} suggested that Cha I is composed of two subclusters with different star 
formation histories, that may have followed independent dynamical evolutions.
Starting from this assumption we estimated $Q$ for two independent samples including the
stars within the two red circles represented in Fig. \ref{fig:map}, which are likely to include 
only stars belonging to one of the two subclusters. For both subclusters we found a value of $Q= 0.76\pm0.06$,
consistent with a randomly distributed sample. This result suggests that if the two subclusters are independent
their dynamical evolution already erased primordial substructures. However, simulations
suggest that the $Q$-parameter is statistically robust only for clusters with a number of stars larger than 100 
(e.g., \citealt{Cartwright:2004, Parker:2015}), while the regions delimited in Fig. \ref{fig:map} include
only 48 and 63 stars. 

Furthermore, we derived the $Q$-parameter ($Q=0.56 \pm 0.06$) from the positions of 60 pre-stellar cores 
found by \cite{Belloche:2011} with a submillimeter survey. The agreement between the value of $Q$ measured for the
cores and for the stars in the smaller ellipse supports the explanation at point a). However, 
as discussed above the number of objects is too low to consider this measurement statistically robust.

\begin{figure}[htb]
   \centering
   \includegraphics[width=8. cm]{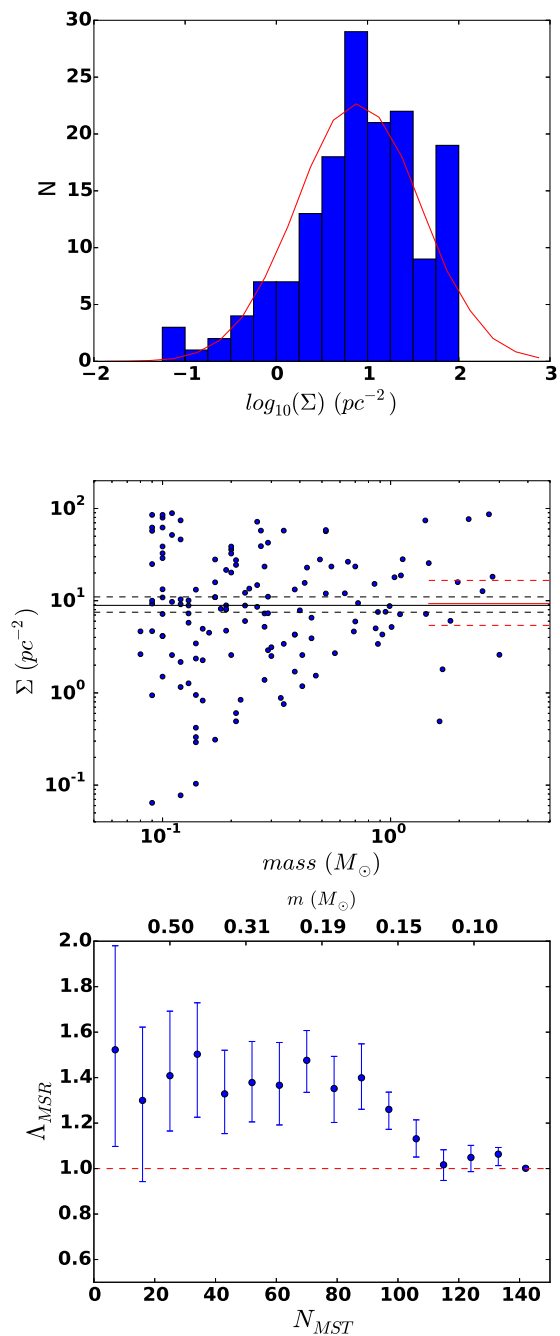}
      \caption{The top panel reports the distribution of the surface density $\Sigma$ and the 
      best-fit model with a lognormal function (continuous red line).
      The middle panel shows the stellar density as defined in Eq. (\ref{eqn:sigma}) as a function of the star mass. 
      The black and the red lines indicate the
      median density of all the stellar sample and the ten most massive stars,  
      respectively. The continuous lines represent the best value, while the dashed lines
      represent their error bars.
      The bottom panel shows the evolution of the mass segregation ratio, $\rm \Lambda_{MSR}$, for the $\rm N_{MST}$
      most massive stars. The top x-axis indicates the lowest mass star $\rm m_L$ within $\rm N_{MST}$ and the
      dashed line corresponds to $\rm \Lambda_{MSR}=1$, i.e., no mass segregation.   }
     \label{fig:fig3}
\end{figure}

\subsection{Stellar density \label{sect:stellar_density}}

The stellar density is a key parameter to derive the effects of the the environment on the evolution of
the star-disc systems and the dynamical status of the clusters. To derive the surface density $\Sigma$ 
we used the same definition as \cite{Bressert:2010}:

\begin{equation}
\Sigma=\frac{N-1}{\pi D_N^2}
\label{eqn:sigma}
\end{equation}

\noindent where $N$ is the $Nth$ nearest neighbour and $D_N$ is the projected distance to that
neighbour. For our calculation we set $N=7$ as in \cite{Bressert:2010}. The top panel in Fig. 
\ref{fig:fig3} shows the density distribution of the stars used to calculate the $Q$ parameter
with the best fit of the distribution with a log-normal function 
(peak $\sim 8~\rm stars~pc^{-2}$ and dispersion $\sigma_{log10\Sigma}=0.67$).
The profile of the distribution is very similar to what observed by \cite{Bressert:2010} 
(see Fig. 1 of the paper) and is well described by a log-normal function in the low-density tail, while
at high density the observed distribution decreases faster than a log-normal function. The reason 
for this deviation from the log-normal model is not clear. However a similar deviation is also
observed in the much larger sample analysed by \cite{Bressert:2010}.
The peak of the distribution is located at lower densities
and the dispersion is lower with respect to the results found by \cite{Bressert:2010}, but this is not surprising
since $\sim$70\% of the young stellar objects used for their analysis belong to the Orion 
star forming region, which is more massive and much denser than Cha I.

Simulations describing the dynamical evolution of young star clusters suggest that the stellar density may depend
on stellar mass, i.e. the density of stars near massive objects can be higher because massive stars act as a potential
well and trap low mass stars \citep[e.g.,][]{Parker:2014}.
The relation between density and mass is plotted in the middle panel
of Fig. \ref{fig:fig3}, which shows that the density around the most massive stars 
(median density $\Sigma = 9.3^{+7.3}_{-3.9}~\rm pc^{-2}$) is consistent within errors
with the surface density for the rest of the stars 
(median density  $\Sigma = 8.9^{+2.1}_{-1.4}\rm pc^{-2}$) \footnote{The best values and the error bars of the 
densities have been calculated by generating 2000 bootstrap resamples. Namely, the best value is the median of the
bootstrap distribution, while the lower and upper values defined by error bars correspond to the 15$^{th}$ and 85$^{th}$
percentiles.}.
This result proves that either the cluster did not go through a sufficient 
dynamical evolution to determine an increase of density around the most massive stars or that these stars are not 
massive enough to act as a potential well and attract low-mass stars.
  
\subsection{Mass segregation \label{sect:mass_seg}}

The last structural property to analyse is the amount of mass segregation. In mass-segregated clusters,
the more massive stars are concentrated in a smaller volume (or projected area on the line of sight)
than lower mass stars. To estimate the level of mass segregation we used the method
introduced by \cite{Allison:2009} and based on the mass segregation ratio $\Lambda_{MSR}$:

\begin{equation}
\Lambda_{MSR}=\frac{\langle l_{average}\rangle}{l_{subset}}^{+\sigma_{5/6}/l_{subset}}_{-\sigma_{5/6}/l_{subset}}
\label{eqn:mass_seg}
\end{equation}

\noindent where $l_{subset}$ is the length of the MST of a subset of stars 
composed of the number $N_{MST}$ most massive stars of the cluster, and $l_{average}$ is the average of the lengths
 of the MSTs of 50 different subsets composed of a number $N_{MST}$ random stars. 
 If $\Lambda_{MSR} > 1$ the MST of the more massive stars is smaller than the MST of a random sample 
so the cluster is mass-segregated, otherwise if $\Lambda_{MSR} < 1$ it is inversely mass segregated 
(i. e., the most massive stars are spread over a larger area than other stars). For estimating the uncertainties
on this ratio, we used the same method as in \cite{Parker:2012b}, namely we considered as lower (upper) error the
length of the MST, which lies at 1/6(5/6) of an ordered list including all the MSTs of the random subsets used to calculate $l_{average}$.
In the bottom panel of Fig. \ref{fig:fig3} we show the evolution of $\Lambda_{MSR}$ as function of 
$N_{MST}$. The upper x-axis shows the smallest mass within the sample of $N_{MST}$ stars.
The plot shows only a marginal evidence of mass segregation, which is not significant since the value of 
$\Lambda_{MSR}$ at higher masses is consistent with $\Lambda_{MSR}$=1, and $\Lambda_{MSR}$ for intermediate mass stars
is above 1 by less than 2-3 error bars, which as estimated by simulations performed by \cite{Parker:2015a} means a 
significance lower than 95\%.

\begin{figure}
   \centering
   \includegraphics[width=9 cm]{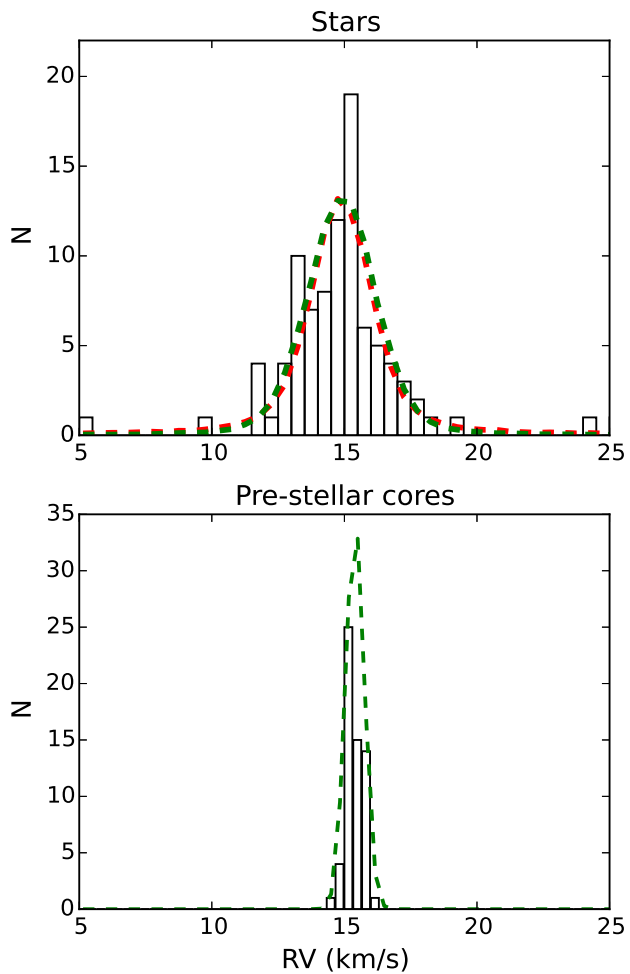}
      \caption{The top panel shows the RV distribution of the full sample of cluster members
      observed by GES. 
      The red and green dashed lines describe the best fit models with a gaussian broadened by 
      the measurement errors and the velocity offsets due to binaries, assuming a fixed binary fraction ($f_{bin}=0.5$) and 
      letting the binary fraction free to vary, respectively. The bottom panel shows the distribution of 
      the RVs of the pre-stellar cores measured by \cite{Tsitali:2015} from the $\rm C^{18}O ~(2-1)$ molecular 
      transition, with overlying the best fit models with a gaussian function.    }
         \label{fig:rvhisto}
\end{figure}

\section{Kinematical properties \label{sect:dyn}}

The precision of the RVs derived from the GES spectra \citep{Jackson:2015} 
allows us to study the kinematical properties of the cluster. We will use the RVs to determine
its global RV dispersion $\sigma_{c}$, to investigate the presence of a RV gradient and to understand, if the 
two different populations identified by \cite{Luhman:2008} have different kinematical properties. For this
analysis, we will use only members of the cluster observed by GES and reported in Table \ref{tab:members},
since only for these stars we have precise measurements of the RV with a proper evaluation of the errors.

\begin{table*}
\caption{Parameters obtained from the fits of the RV distributions with 1$\sigma$ errors \label{tab:RVfit}.}
\centering
\begin{tabular}{ccccccc}
\hline\hline
     &  $v_{c}$  & $\sigma_{c}$ & $f_{bin}$ &  $\alpha$ & $\beta$  & $\ln(L_{max})$\\
     &  $\rm (km~s^{-1})$  & $\rm (km~s^{-1})$ &  & $\rm (km~s^{-1}~deg^{-1})$ & $\rm (km~s^{-1}~deg^{-1})$ & \\
\hline
Cha I (fit 1) & 14.88$\pm$0.15 & 0.94$\pm$0.15 & 0.5           & 0              & 0              & -183.68\\
Cha I (fit 2) & 14.90$\pm$0.15 & 1.11$\pm$0.11 & 0.18$\pm$0.11 & 0              & 0              & -181.52 \\
Cha I (fit 3) & 14.87$\pm$0.15 & 1.08$\pm$0.14 & 0.17$\pm$0.11 & -0.21$\pm$0.13 & 0.11$\pm$0.24  & -180.16\\
Cha I N      & 15.29$\pm$0.22 & 0.95$\pm$0.18 & 0.18           &  0             & 0              & -51.94 \\
Cha I S      & 14.36$\pm$0.22 & 0.87$\pm$0.24 & 0.18           &  0             & 0              & -78.06 \\
Cha I Outer  & 14.98$\pm$0.30 & 1.17$\pm$0.28 & 0.18           &  0             & 0              & -47.05 \\
\hline
\end{tabular}
\end{table*}

\subsection{Radial Velocity dispersion}

The RV distribution of the cluster members is shown 
in the top panel of Fig. \ref{fig:rvhisto}. We modeled the distribution using a maximum-likelihood method developed by 
\cite{Cottaar:2012} and already used in several works \citep[e.g.,][]{Jeffries:2014, Foster:2015, Sacco:2015}, which
allows us to properly take into account the errors on each star and the presence of binaries.
Specifically, we assume that the stellar RVs have an intrinsic Gaussian distribution (with mean $v_{c}$ and 
standard deviation $\sigma_{c}$) broadened by the measurement uncertainties and the velocity offsets due to binary orbital motion.
The distribution of the offsets is calculated numerically
by a code\footnote{Available online at https://github.com/MichielCottaar/velbin.} developed by \cite{Cottaar:2012}, 
considering three different assumptions: 
a) binary periods follow a log-normal distribution with mean period 5.03 and dispersion 2.28 in $log_{10}$ days 
\citep{Raghavan:2010}; b) the secondary to primary ratio ($q$) follows a power-law
$\frac{dN}{dq}\sim q^{-0.5}$ for 0.1 < q < 1 \citep{Reggiani:2011}; c) the distribution 
of eccentricities is flat between 0 and a maximum value which depends on the period according 
to Eqn. 6 from \cite{Parker:2009}.

\begin{figure}
   \centering
   \includegraphics[width=9 cm]{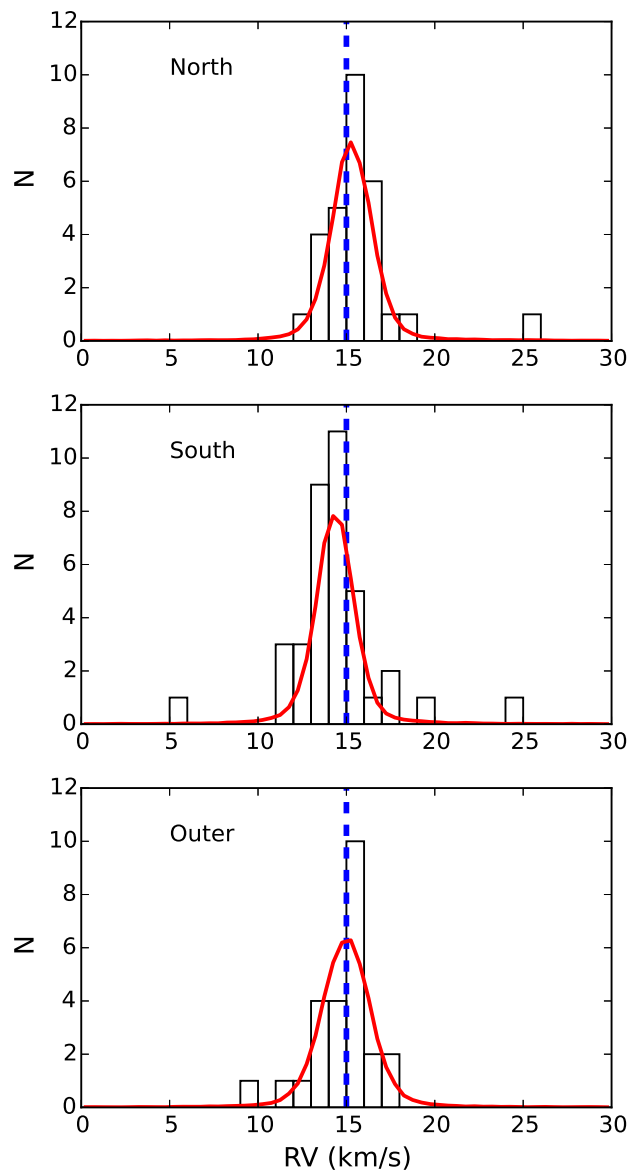}
      \caption{The three panels from the top to the bottom show the RV distributions of the North, the South subclusters,
      and the stars dispersed in the outer regions, respectively. The dashed blue line marks the velocity of 
      15 $\rm km~s^{-1}$ and the red line is the best-fit the distribution with the same model used for the full sample.}
         \label{fig:rvhisto_sub}
\end{figure}

We performed two fits. In the first one (fit 1 in Table \ref{tab:RVfit}), we kept the fraction of binaries fixed at 
$f_{bin}=0.5$, while in the second one (fit 2 in Table \ref{tab:RVfit}), it was left free to vary. 
In both fits, we consider only stars in the range $0 < RV < 30~\rm km~s^{-1}$,
since stars outside this range are either binaries or stars with a miscalculated RV due to the presence of
strong emission lines. The parameters derived by the two fits are reported in the first two rows of Table \ref{tab:RVfit}
and the best fit functions are plotted in Fig. \ref{fig:rvhisto}. Since the two models are nested,
to evaluate the parameters of which fit to adopt we can perform a likelihood-ratio test. This gives a 
probability P($L_{fit1}/L_{fit2}$) = 3.8 \%, which indicates, with a marginal level of significance, that
the fit with a fixed binary fraction can be rejected and the parameters from the second fit can be adopted.

Our results are in agreement with previous estimates of the central cluster velocity and of the velocity dispersion from 
\cite{Joergens:2006} ($v_{c}= 14.7~\rm km~s^{-1}$ and $\sigma_{c}= 1.3~\rm km~s^{-1}$).

\subsection{Radial velocity gradient}

To investigate the presence of a RV gradient in the stellar population, we fitted the RV distribution with the same function
discussed in the previous section, but instead of considering the mean cluster velocity $v_{c}$ as a single
free parameter, we assumed that the velocity $v_{c}=v_{c0}+\alpha\Delta RA+\beta\Delta DEC$, where
$\Delta RA$ and $\Delta DEC$ are the RA and DEC shifts of each star with respect to a fixed position calculated
as the median of the star positions, and $v_{c0}$, $\alpha$ and $\beta$ are free parameters of the fit together
with $\sigma_{c}$, which is assumed to be constant over the whole region.
The result of this fit is reported in Table \ref{tab:RVfit} (fit 3). The parameters $v_{c0}$ and $\sigma_{c}$ 
are in agreement with the results found with the previous fits and the components of the RV gradient $\alpha$ and $\beta$
are consistent within two standard deviations with zero. Since the function used for fit2 is the same as 
fit 3, when we fix the parameters $\alpha$ and $\beta$ to zero, we can use the likelihood-ratio test
to compare the model with and without a gradient. The probability  P($L_{max}(fit2)/L_{max}(fit3)$)= 26\%, 
so we conclude that there is no evidence of the presence of a RV gradient in the cluster.

\subsection{Kinematical properties of the sub-clusters \label{sect:subclusters}}

\cite{Luhman:2007} suggested that Cha I is composed of two subclusters with 
different star formation histories.
To understand if these two populations have different kinematical properties we divided 
our sample in three groups: 
1) one composed of 29 stars located within the upper circle drawn in Fig.\ref{fig:map} 
with a blue dashed line, which approximately 
defines the boundary of the northern part of the cloud; 2) one composed of 37 stars 
within the lower circle, which defines the boundary of the southern cloud; 
3) one composed of 25 stars located in the outer regions..

The RV distributions of the three samples are shown in Fig. \ref{fig:rvhisto_sub} and the results 
from the fits of the distributions are reported in the last rows of Table \ref{tab:RVfit}. 
The central RVs of the two clusters concentrated around the clouds differ by $\sim 1~\rm km~s^{-1}$ at 
2$\sigma$ level of significance and on the basis of a Kolmogorov-Smirnov test the probability 
that the two distributions are
part of the same population is $<$1\%. The kinematical properties of the stars located 
in the outer regions are closer to what found for the northern stars, suggesting that the 
majority of the outer stars belong to the northern cluster. This is consistent with the 
hypothesis suggested by \cite{Luhman:2007}, that the northern cluster started 
to form earlier and, therefore, it is going through a more advanced stage of its evolution.
\cite{Lopez-Marti:2013a} already tried to kinematically separate the two subclusters
using proper motions, finding no evidence of different velocities. They do not report any 
upper limit on the velocity separation between subclusters, so it is difficult to compare their 
data with our result. However, the precision of proper motions used for their work is 
lower than RVs from the Gaia-ESO Survey.

\begin{figure*}
   \centering
\includegraphics[width =16.6 cm]{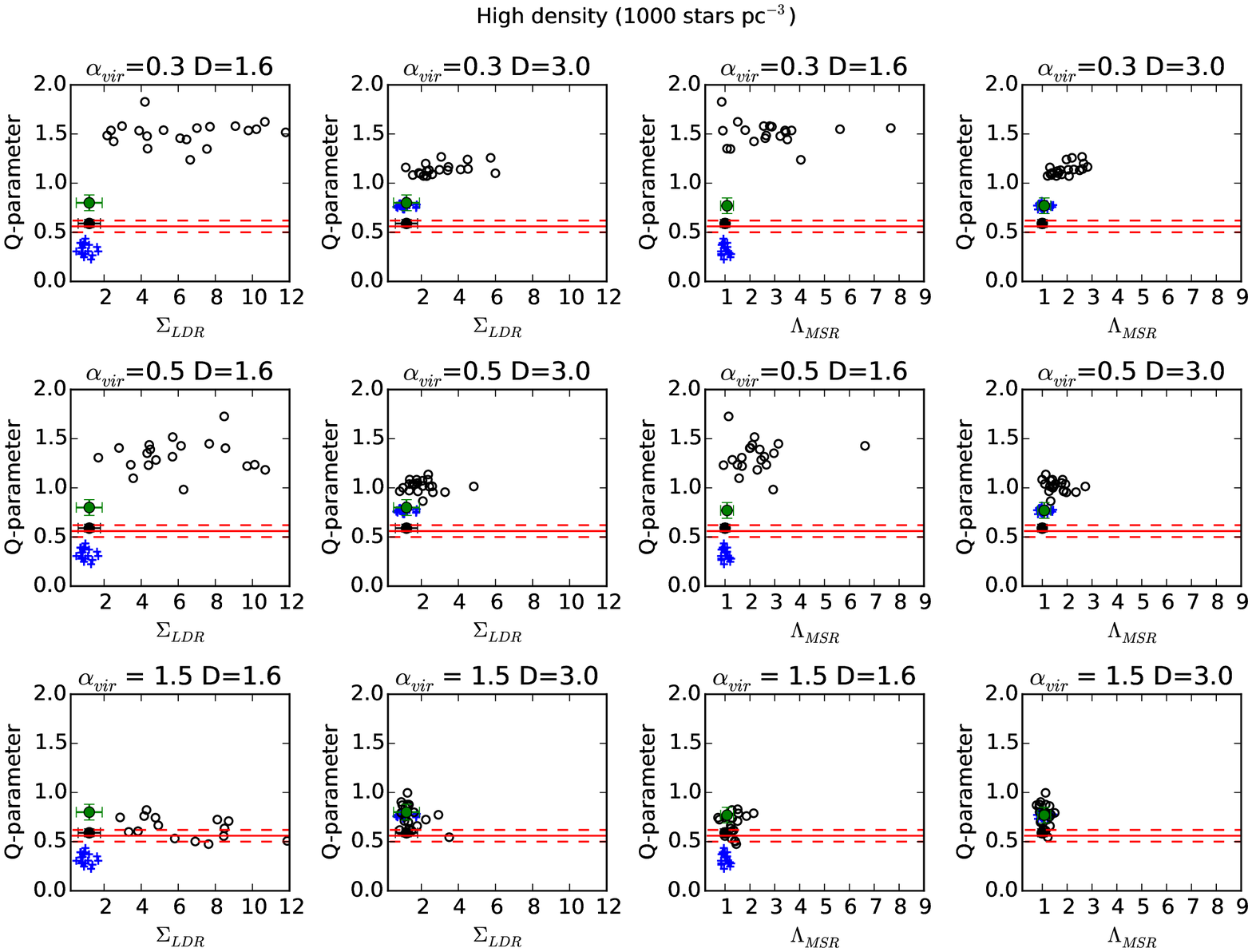}  
  \includegraphics[width =16.6 cm]{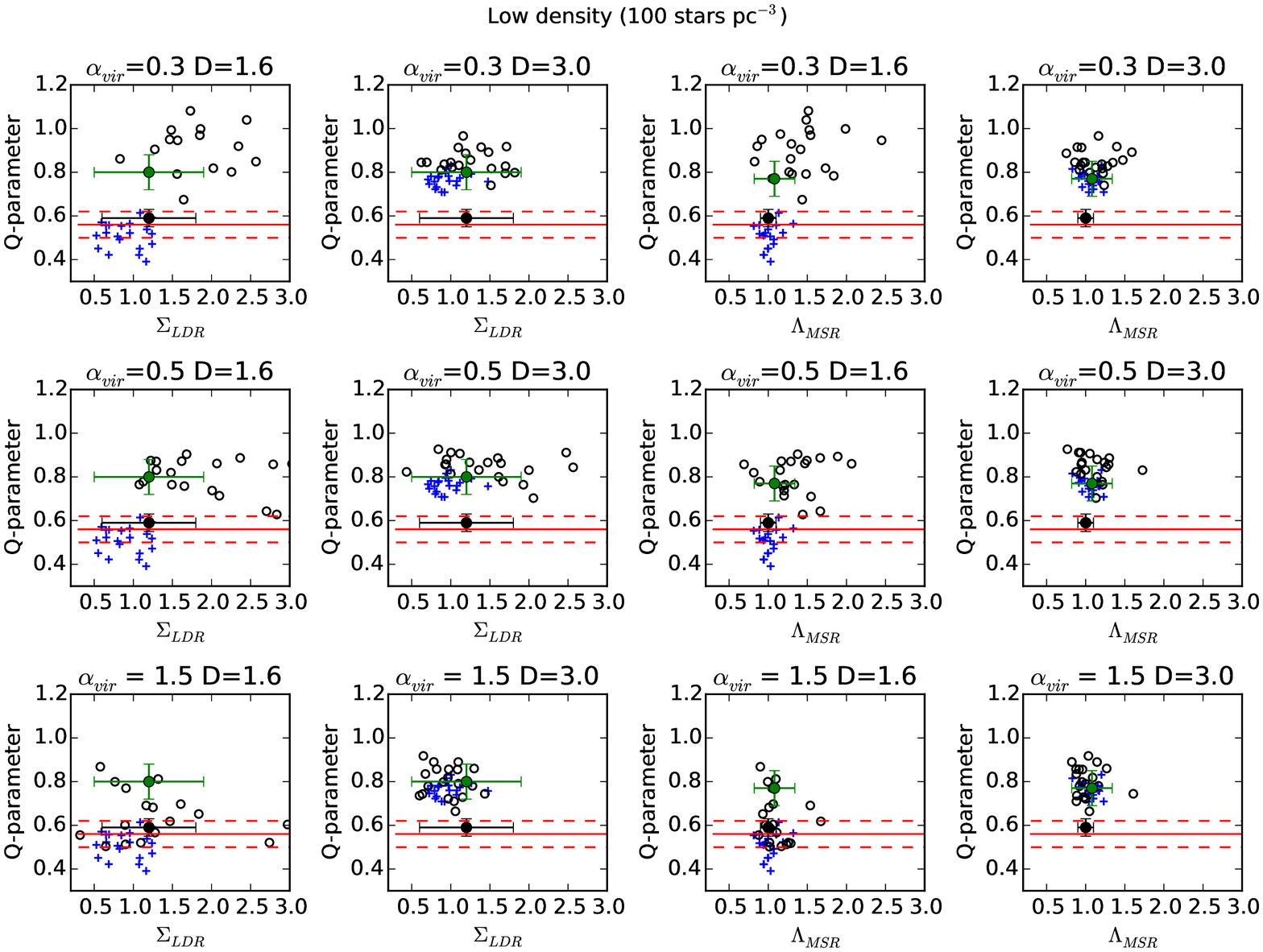}  
      \caption{Comparison between the observed structural properties of Cha I and simulated
       clusters with high (n$_{stars}$ = 1000 stars pc$^{-3}$, top panels) and low (n$_{stars}$ = 100 stars pc$^{-3}$, bottom panels) 
       initial stellar density from \cite{Parker:2014}.
      The panels on the left and on the right show the $Q$-parameter as function of $\Sigma_{LDR}$ 
      (i.e., the ratio between the median superficial density of the most massive stars and the rest of the sample), 
      and  $\Lambda_{MSR}$ (for the ten most massive stars), respectively. 
      The simulations differ for the initial virial ratio $\alpha_{vir}$ and the initial level of substructure 
      (D = 1.6 is a highly substructured cluster and D = 3.0 is a roughly uniform sphere). 
      Blue crosses and black circles represent the simulated clusters at the initial conditions and after 2 Myr 
      evolution, respectively. Green and black dots represent the properties of Cha I for the full sample of members and 
      only for the stars in the embedded region within the smallest ellipse in Fig. \ref{fig:mst}, respectively. 
      The red lines trace the $Q$-parameter estimated for the pre-stellar cores with errors.}
         \label{fig:parker_models}
\end{figure*}

\section{Discussion}

The main goal of this work is to study the physical processes leading to the formation and the dynamical evolution
of small star clusters. In the next sections, we will compare the structural and dynamical 
properties of the stellar populations in Cha I with the properties of pre-stellar 
cores and with some numerical models 
describing the early stages of the star cluster evolution.

\subsection{Structural properties}

Using N-body simulations, \cite{Parker:2014} studied the evolution of the level of substructure and the
mass segregation in young star clusters.
In Fig. \ref{fig:parker_models} we compare the results of the simulations performed by \cite{Parker:2014} for 
clusters with high ($\rm n_{stars}\sim1000~stars~pc^{-1}$) and low ($\rm n_{stars}\sim100~stars~pc^{-1}$) stellar
density with the structural properties of Cha I derived in Sect. \ref{sect:struct}. The simulations differ for 
the initial virial ratio ($\alpha_{vir}$) and the initial fractal dimension $D$, which indicated the level 
of substructure (D = 1.6 is a highly substructured cluster and D = 3.0 is a roughly uniform sphere). 
The figure shows the initial conditions of the simulated clusters (t = 0 Myr) and their status after 2 Myr.

None of the simulated clusters with high stellar density reproduce the structural properties of Cha I, 
with the exception of the case of a supervirial cluster with no substructure. However, even in this case, 
the properties of the simulated clusters at the initial conditions are not consistent with the properties of the 
pre-stellar cores.
This results is not surprising, since the stellar density of the simulated clusters is much higher than the observed 
density in Cha I of both stars and pre-stellar cores, and further supports the hypothesis that 
Cha I did not form in a high density environment, in contrast to the hypothesis advanced 
by \cite{Marks:2012}.

The properties of the low-density simulated clusters are much closer to the properties of Cha I.
In particular, virial and supervirial simulated clusters are consistent with the overall properties of the
cluster after 2 Myr of dynamical evolution. Furthermore, the simulations with a high level of
substructure at t = 0 Myr are consistent with the properties of embedded stars and pre-stellar cores, if 
we assume that these represent the properties of the cluster at its formation. Otherwise, 
according to the simulations, for a cluster which is initially sub-virial we should observe 
a level of mass segregation after 2 Myr, which we do not observe in Cha I. 

To summarize, according to this analysis, Cha I formed in a low-density environment with a virial
ratio $\alpha_{vir}\geq 0.5$ and a high level of subtructures. It has erased 
substructure due to dynamical interactions and will likely disperse
 in the Galactic field. 
 A similar scenario has been proposed for the more evolved cluster Gamma Velorum \citep{Jeffries:2014, Mapelli:2015, 
 Sacco:2015}. It would be interesting to perform a direct measurement of the virial ratio in Cha I. However, due to the 
 highly asymmetric structure of both the stellar and the gas component of the cluster it is difficult to estimate the 
 virial ratio without any information about its structure along the line of sight. This information will be provided by the astrometric mission 
{\it Gaia} for most of the optically visible stars.

It is worth mentioning a few caveats concerning the comparisons between the simulations 
performed by \cite{Parker:2014} and our results: a) as proven by the large area of the 
parameter space covered by the simulations in each panels of Fig. \ref{fig:parker_models},
N-body simulations, especially of low-density clusters, are partially degenerate
namely, the same initial conditions may lead to clusters with very different properties in the
$Q$ vs. $\Sigma_{LDR}$ and $Q$ vs. $\Lambda_{MSR}$ plots. In particular, the comparison between
our results and models give strong constrains about the initial density of the cluster, but less
stronger constrains about other critical properties, like the initial virial ratio. The analysis of 
other young star clusters similar to Cha I and the definition of new diagnostics of the dynamical status
of star clusters using also kinematic data can help to overcome this limitation; b) 
N-body simulations do not include the presence of gas, which in the case of Cha I is the 
main component of the potential energy of the systems.
The effects of the gas in the evolution of star clusters is a very debated topic. Some authors
(e.g., \citealt{Kruijssen:2012}) suggest that the influence of the gas in the cluster dynamics
is negligible, but a direct comparison between simulations describing in a consistent way the evolution of
gas and stars is required to provide final answers to this issue.; 
c) the simulations discussed in this paper assume that all the stars are coeval, while the age spread in Cha I is 
larger than its median age. The origin of the age spread in young clusters is not
clear and is not reproduced by any of the state-of-the-art simulations. More sophisticated simulations
and precise and complete data are required to fully understand the star formation history of clusters
like Cha I.

\subsection{Kinematical properties}

As shown in Fig. \ref{fig:rvhisto}, the most clear result of the kinematical analysis is the large discrepancy between the velocity dispersion of the stars 
($\rm \sigma_{stars} = 1.10 \pm 0.15~km~s^{-1}$) and that of pre-stellar cores ($\rm \sigma_{cores}\sim0.3~km~s^{-1}$) 
derived from submillimeter observations of molecular transitions by \cite{Tsitali:2015}. As noted in Tab. \ref{tab:RVfit}, 
the velocity dispersion of the stellar component does not depend on the sample of stars used for the fit. In fact, the two sub-clusters around 
the molecular cloud and the sample of stars located in the outer region have similar velocity dispersions, which are in all cases much higher 
than the dispersion measured for the pre-stellar cores. 
A similar discrepancy between the pre-stellar cores and the stars has been observed in the $\rho$ Oph star forming region, 
in the young cluster NGC 1333, and in Orion. The velocity dispersion of the stellar component 
in $\rho$ Oph ($\rm \sigma_{stars} = 1.14\pm 0.35~km~s^{-1}$) 
was derived from the Gaia-ESO observations of the optically visible stars around 
the main cloud L1688 by \cite{Rigliaco:2016}, who suggested
that the cluster is bound and in virial equilibrium, while the velocity dispersion of the 
cores ($\rm \sigma_{cores}\sim 0.4~km~s^{-1}$) was 
estimated by \cite{Andre:2007}, who suggested that the cores are subvirial. 
The kinematical properties of both the cores and the stars 
of NGC 1333 have been analysed by \cite{Foster:2015}, who also found that the stars are virial 
($\rm \sigma_{stars} = 0.92 \pm 0.12~km~s^{-1}$),
while the cores ($\rm \sigma_{cores}\sim 0.5~km~s^{-1}$) are sub-virial. 
They suggested that the discrepancy between stars and cores can be due either
to the magnetic field having a strong influence on the cores and/or to the global collapse of 
the cluster after the protostellar phase.
A similar conclusion has been obtained by N-body simulations carried out by \cite{Parker:2016}, 
who found that clusters starting as subvirial
undergo cool collapse, so the dynamical interaction among stars quickly inflate the distribution. 
However, for small low-density
clusters \cite{Parker:2016} found a lower velocity dispersion ($\rm \sigma\sim 0.5~km~s^{-1}$) 
than observed. This discrepancy could be 
associated to the lack of gas in the N-body simulations, since the presence of a significant amount 
of gas reduces the virial ratio and leads to the collapse of clusters with higher velocity 
dispersion than in the case without gas (\citealt{Leigh:2014}, Mapelli et al., in prep.).
The morphology and the kinematic of gas, protostars and pre-main sequence stars has been studied 
in Orion A by \cite{Stutz:2016}. They propose that protostars are ejected from the filaments due 
to magnetically induced transverse waves. This slingshot-like mechanism is responsible of the velocity discrepancy
between young stars and protostars still within the filaments.

The second result of our kinematical analysis is the discrepancy ($\rm \sim 1~km~s^{-1}$) between the 
central velocities of the two sub-clusters located around the northern and southern clouds. 
This is not surprising, since \cite{Luhman:2007} suggested that Cha I is composed of two 
components with different star formation histories. Furthermore, recent studies show that multiple populations 
(e.g., \citealt{Jeffries:2014, Sacco:2015}) and RV gradients (e.g., \citealt{Tobin:2015}) are 
common in young clusters and star 
forming regions. According to the submillimeter observations, the mean velocities of the cores in 
the north and the south clusters also
differ by $\rm \sim0.3~km~s^{-1}$. However, the discrepancy is in the opposite direction with 
respect to what we found for the stars, namely
the cores in the south have a higher redshift than in the north. The reason of this anti-correlation 
between stars and cores is not clear,
but this result supports a scenario where the dynamics of the cores is independent 
from the dynamics of the stellar populations.

\section{Conclusions}

In this work we present a new analysis of the spectroscopic parameters derived from the Gaia-ESO Survey 
observations of the young cluster Cha I aimed at investigating the structural and dynamical properties of the cluster. 
We obtained the following main results.

\begin{enumerate}

\item An evident discrepancy between the 
velocity dispersion of the stellar component ($\rm \sigma_{star} = 1.10\pm0.15~km~s^{-1}$) derived 
by Gaia-ESO spectra and the dispersion ($\rm \sigma_{cores} = 0.3~km~s^{-1}$) of pre-stellar cores 
derived by submillimeter observations. A similar discrepancy
has observed in the young embedded clusters $\rho$ Oph, NGC 1333 and in Orion. 
The origin of such a large discrepancy is not clear. It could be related to the effect of the magnetic
field on the protostars or the filaments where they form, or to two-body stellar dynamical interactions
following the cluster formation. We will further investigate this issue in a forthcoming paper 
(Mapelli et al. in prep.).

\item Analysing independently the RV distributions of the two sub-clusters located around the two main molecular clouds, we found that the central RVs differ
by $\rm \sim 1~km~s^{-1}$. This result supports the evidence found by \cite{Luhman:2007} that Cha I is 
composed of two sub-clusters with different star formation
histories and a scenario where young clusters do not form as monolithic systems but from the merging of smaller subsystems.

\item A new membership analysis based on three independent spectroscopic criteria led to the confirmation 
of all the previously known members, for which new astrophysical parameters from the Gaia-ESO Survey
are available, and to the discovery of six new members in Cha I and one new member of the $\epsilon$ Cha association, 
which are all located in the outer part of the cluster.

\item The level of substructure of the cluster measured using the $Q$-parameter 
defined by \cite{Cartwright:2004} depends on the sample used for the calculation. 
If we consider only the stars in the inner region the value of $Q$ indicates that the cluster 
is highly substructured, while if we take into account the full sample of members the spatial
 distribution of the cluster is consistent with a random sample. It is not clear if this 
 trend has a physical origin or if it is the result of a bias due to differential extinction 
 in the inner region of the cluster or incomplete target selection in the outer region.

\item As observed in other low-mass young star clusters, Cha I is not mass-segregated and 
its superficial density follows a log-normal distribution, with the exception of its high mass
end, which follows a steeper trend.

\item The comparison between the observed structural properties of Cha I and the results of N-body simulations performed by
\cite{Parker:2014} suggests that the cluster formed as highly substructured, and virial or supervirial. 
However, discrepancies between the simulated clusters and Cha I (e.g., the lack of gas in the simulated clusters) 
may affect this comparison.

\end{enumerate}

\input{Tab1_elec_mod.tex}
\input{Tab2_elec_bis.tex}
\begin{acknowledgements}
These data products have 
been processed by the Cambridge Astronomy Survey Unit (CASU) at the Institute of Astronomy, University of Cambridge, and by the FLAMES/UVES reduction team at 
INAF/Osservatorio Astrofisico di Arcetri. These data have been obtained from the Gaia-ESO Survey Data Archive, prepared and hosted by the Wide Field Astronomy 
Unit, Institute for Astronomy, University of Edinburgh, which is funded by the UK Science and Technology Facilities Council.
This work was partly supported by the European Union FP7 programme through ERC grant number 320360 and by the Leverhulme Trust through grant RPG-2012-541. 
We acknowledge the support from INAF and Ministero dell' Istruzione, dell' Universit\`a' e della Ricerca (MIUR) in the form of the grant "Premiale VLT 2012". 
This work has been partially supported by PRIN-INAF-2014.
The results presented here benefit from discussions held during the Gaia-ESO workshops and conferences supported by the ESF (European Science Foundation) through 
the GREAT Research Network Programme.
MM acknowledges financial support from the Italian Ministry of Education, University and Research (MIUR) through grant FIRB 2012 RBFR12PM1F, from INAF through grant PRIN-2014-14, and from the MERAC Foundation. 
\end{acknowledgements}


\bibliographystyle{aa}
\bibliography{/Users/sacco/LETTERATURA/bibtex_all}

\end{document}

%% file: Tab1.tex
\begin{table*}[ht]
\caption{Members of the cluster observed by the Gaia-ESO Survey \label{tab:members}}
\begin{tabular}{cccccccccc}
\hline \hline
Cname & RA & DEC & RV\tablefootmark{a} & $T_{eff}$ & $\gamma$\tablefootmark{b} & EW(Li) & $H\alpha$10\%\tablefootmark{c} & Inst.\tablefootmark{d} & memb\tablefootmark{e} \\
 & (J2000) & (J2000) & $\rm (km~s^{-1})$ & (K) &  & (m\AA) & $\rm (km~s^{-1})$ &  &  \\
\hline
10550964-7730540 & 163.79017 & -77.51500 & 16.83$\pm$0.90 & - & - & 725$\pm$23 & 128$\pm$5 & G & Y \\
10555973-7724399 & 163.99887 & -77.41108 & - & 3640$\pm$300 & - & 120$\pm$58 & 441$\pm$17 & U & Y \\
10561638-7630530 & 164.06825 & -76.51472 & 12.56$\pm$2.01 & - & - & - & 215$\pm$5 & G & Y \\
10563044-7711393 & 164.12683 & -77.19425 & 15.69$\pm$0.27 & 4351$\pm$505 & 0.968$\pm$0.005 & 300$\pm$17 & 390$\pm$8 & G & Y \\
10563146-7618334\tablefootmark{f} & 164.13108 & -76.30928 & 15.41$\pm$0.54 & 3319$\pm$48 & - & 597$\pm$11 & 142$\pm$4 & G & N \\
10574219-7659356 & 164.42579 & -76.99322 & 16.14$\pm$0.26 & 3452$\pm$183 & 0.895$\pm$0.004 & 571$\pm$16 & 272$\pm$6 & G & Y \\
10575376-7724495 & 164.47400 & -77.41375 & 15.71$\pm$0.32 & 3426$\pm$88 & 0.857$\pm$0.013 & 632$\pm$10 & 138$\pm$8 & G & N \\
10590108-7722407 & 164.75450 & -77.37797 & 15.64$\pm$0.40 & 4135$\pm$125 & - & 375$\pm$19 & 375$\pm$8 & U & Y \\
10590699-7701404 & 164.77912 & -77.02789 & 17.93$\pm$0.26 & 4981$\pm$260 & - & 390$\pm$1 & 440$\pm$9 & G & Y \\
11004022-7619280 & 165.16758 & -76.32444 & 15.73$\pm$0.42 & - & - & 584$\pm$18 & 230$\pm$5 & G & Y \\
\hline
\end{tabular}
\tablefoot{A full version of the table is available at the CDS.}
\tablefoottext{a}{For spectra with a signal-to-noise ratio lower than three we did not report any velocity.} \\ 
\tablefoottext{b}{Empirical gravity indicator defined by \cite{Damiani:2014}.}\\ 
\tablefoottext{c}{Width at 10\% of the peak of the H$\alpha$ line.}\\ 
\tablefoottext{d}{The letters "G" and "U" indicate GIRAFFE and UVES, respectively.}\\
\tablefoottext{e}{The letters "Y" and "N" indicate the star is a known member or not, respectively}\\
\tablefoottext{f}{This star is likely a member of the $\epsilon$ Cha association.}
\end{table*}

%% file: Tab2.tex
\begin{table}[ht]
\caption{Known cluster members from the literature and new members used to study the structural properties of the cluster  \label{tab:struct}}
\begin{tabular}{cccccc}
\hline \hline
RA & DEC & $T_{eff}$ & $Log\left (\frac{L_{Bol}}{L_{\sun}}\right )$ & prov\tablefootmark{a} & Mass \\
(J2000) & (J2000) & (K) &  &  & $(M_{\sun})$ \\
\hline
161.65812 & -77.60097 & 7200 & 0.95 & L & 1.64 \\
163.15392 & -74.67464 & 3161 & -1.00 & L & 0.14 \\
163.41575 & -77.20939 & 3451 & -1.49 & L & 0.30 \\
163.79017 & -77.51500 & 3198 & -1.08 & L & 0.15 \\
163.99887 & -77.41108 & 3640 & -0.36 & G & 0.37 \\
163.99887 & -77.41108 & 3640 & -0.36 & G & 0.37 \\
164.06825 & -76.51472 & 3044 & -1.51 & L & 0.09 \\
164.12683 & -77.19425 & 4350 & 0.06 & G & 0.85 \\
164.42579 & -76.99322 & 3451 & -0.33 & G & 0.28 \\
\hline
\end{tabular}
\tablefoot{A full version of the table is available at the CDS.\\ \
\tablefoottext{a}{The letters G and L indicate that the data used for deriving the stellar mass are retrieved from the \
literature and the GES archive, respectively.}}
\end{table}

%% file: Tab1_elec_mod.tex
\addtocounter{table}{-4}
\onllongtab{
\begin{longtable}{ccccccccccc}
\caption{Members of the cluster observed by the Gaia-ESO Survey \label{tab:members_elec}}\\
\hline \hline
Cname & RA & DEC & RV\tablefootmark{a} & $T_{eff}$ & $\gamma$\tablefootmark{b} & EW(Li) & $H\alpha$10\%\tablefootmark{c} & Inst.\tablefootmark{d} & memb\tablefootmark{e} \\
 & (J2000) & (J2000) & $\rm (km~s^{-1})$ & (K) &  & (m\AA) & $\rm (km~s^{-1})$ &  &  \\
\hline
\endfirsthead
\caption{continued.}\\
\hline \hline
Cname & RA & DEC & RV\tablefootmark{a} & $T_{eff}$ & $\gamma$\tablefootmark{b} & EW(Li) & $H\alpha$10\%\tablefootmark{c} & Inst.\tablefootmark{d} & memb\tablefootmark{e} \\
 & (J2000) & (J2000) & $\rm (km~s^{-1})$ & (K) &  & (m\AA) & $\rm (km~s^{-1})$ &  &  \\
\hline
\endhead
\hline
\endfoot
10550964-7730540 & 163.79017 & -77.51500 & 16.83$\pm$0.90 & - & - & 725$\pm$23 & 128$\pm$5 & G & Y \\
10555973-7724399 & 163.99887 & -77.41108 & - & 3640$\pm$300 & - & 120$\pm$58 & 441$\pm$17 & U & Y \\
10561638-7630530 & 164.06825 & -76.51472 & 12.56$\pm$2.01 & - & - & - & 215$\pm$5 & G & Y \\
10563044-7711393 & 164.12683 & -77.19425 & 15.69$\pm$0.27 & 4351$\pm$505 & 0.968$\pm$0.005 & 300$\pm$17 & 390$\pm$8 & G & Y \\
10563146-7618334\tablefootmark{f} & 164.13108 & -76.30928 & 15.41$\pm$0.54 & 3319$\pm$48 & - & 597$\pm$11 & 142$\pm$4 & G & N \\
10574219-7659356 & 164.42579 & -76.99322 & 16.14$\pm$0.26 & 3452$\pm$183 & 0.895$\pm$0.004 & 571$\pm$16 & 272$\pm$6 & G & Y \\
10575376-7724495 & 164.47400 & -77.41375 & 15.71$\pm$0.32 & 3426$\pm$88 & 0.857$\pm$0.013 & 632$\pm$10 & 138$\pm$8 & G & N \\
10590108-7722407 & 164.75450 & -77.37797 & 15.64$\pm$0.40 & 4135$\pm$125 & - & 375$\pm$19 & 375$\pm$8 & U & Y \\
10590699-7701404 & 164.77912 & -77.02789 & 17.93$\pm$0.26 & 4981$\pm$260 & - & 390$\pm$1 & 440$\pm$9 & G & Y \\
11004022-7619280 & 165.16758 & -76.32444 & 15.73$\pm$0.42 & - & - & 584$\pm$18 & 230$\pm$5 & G & Y \\
11011875-7627025 & 165.32812 & -76.45069 & 13.27$\pm$0.26 & 3940$\pm$236 & 0.905$\pm$0.004 & 580$\pm$21 & 175$\pm$11 & G & Y \\
11021927-7536576 & 165.58029 & -75.61600 & 15.31$\pm$0.90 & - & - & 566$\pm$33 & 94$\pm$4 & G & Y \\
11022491-7733357 & 165.60379 & -77.55992 & 14.33$\pm$0.40 & 4519$\pm$174 & - & 471$\pm$12 & 382$\pm$8 & U & Y \\
11023265-7729129 & 165.63604 & -77.48692 & 15.35$\pm$0.31 & 3610$\pm$104 & 0.885$\pm$0.008 & 618$\pm$13 & 148$\pm$10 & G & Y \\
11025504-7721508 & 165.72933 & -77.36411 & 14.97$\pm$0.54 & 2900$\pm$262 & 0.919$\pm$0.013 & 174$\pm$11 & 322$\pm$10 & G & Y \\
11035144-7540335 & 165.96433 & -75.67597 & 9.66$\pm$1.45 & - & - & 443$\pm$73 & 144$\pm$6 & G & N \\
11035682-7721329 & 165.98675 & -77.35914 & 5.32$\pm$0.32 & 3443$\pm$95 & 0.865$\pm$0.011 & 616$\pm$25 & 135$\pm$2 & G & Y \\
11045100-7625240 & 166.21250 & -76.42333 & 13.77$\pm$0.40 & 4571$\pm$334 & - & 576$\pm$17 & 175$\pm$7 & U & Y \\
11045285-7625514 & 166.22021 & -76.43094 & 13.91$\pm$0.26 & 4031$\pm$267 & 0.911$\pm$0.003 & 583$\pm$18 & 120$\pm$3 & G & Y \\
11045701-7715569 & 166.23754 & -77.26581 & -67.23$\pm$5.83 & - & - & - & 372$\pm$48 & G & Y \\
11050752-7812063 & 166.28133 & -78.20175 & 15.61$\pm$1.66 & - & - & 399$\pm$15 & 181$\pm$87 & G & Y \\
11051467-7711290 & 166.31113 & -77.19139 & 17.02$\pm$0.38 & 3433$\pm$200 & 0.892$\pm$0.010 & 525$\pm$10 & 241$\pm$13 & G & Y \\
11052272-7709290 & 166.34467 & -77.15806 & 13.34$\pm$1.57 & - & - & - & 126$\pm$5 & G & Y \\
11052472-7626209 & 166.35300 & -76.43914 & 15.23$\pm$0.28 & 3576$\pm$73 & 0.881$\pm$0.007 & 567$\pm$17 & 104$\pm$3 & G & Y \\
11054300-7726517 & 166.42917 & -77.44769 & 15.29$\pm$0.38 & 3223$\pm$98 & 0.944$\pm$0.007 & 542$\pm$8 & 138$\pm$5 & G & Y \\
11055261-7618255 & 166.46921 & -76.30708 & 15.00$\pm$0.55 & 4406$\pm$740 & - & 609$\pm$13 & 202$\pm$9 & G & Y \\
11055780-7607489 & 166.49083 & -76.13025 & - & - & - & - & - & U & Y \\
11060011-7507252 & 166.50046 & -75.12367 & 13.88$\pm$1.20 & - & - & 504$\pm$17 & 142$\pm$6 & G & Y \\
11062555-7633418 & 166.60646 & -76.56161 & 164.11$\pm$1.61 & - & - & - & 305$\pm$25 & G & Y \\
11064180-7635489 & 166.67417 & -76.59692 & 14.96$\pm$0.51 & 3273$\pm$16 & 0.928$\pm$0.016 & 332$\pm$14 & 416$\pm$10 & G & Y \\
11064510-7727023 & 166.68792 & -77.45064 & 14.69$\pm$0.40 & 4343$\pm$147 & - & 401$\pm$51 & 503$\pm$38 & U & Y \\
11065733-7742106 & 166.73888 & -77.70294 & 12.88$\pm$0.28 & 3436$\pm$62 & 0.901$\pm$0.007 & 606$\pm$20 & 155$\pm$7 & G & Y \\
11065906-7718535 & 166.74608 & -77.31486 & -606.45$\pm$2.79 & - & - & - & 432$\pm$9 & G & Y \\
11070919-7723049 & 166.78829 & -77.38469 & - & - & - & - & - & G & Y \\
11071148-7746394 & 166.79783 & -77.77761 & 17.24$\pm$0.54 & 3709$\pm$133 & - & 640$\pm$8 & 288$\pm$7 & G & Y \\
11071206-7632232 & 166.80025 & -76.53978 & 16.59$\pm$0.25 & 3980$\pm$135 & 0.921$\pm$0.003 & 578$\pm$9 & 523$\pm$10 & G & Y \\
11071330-7743498 & 166.80542 & -77.73050 & 14.46$\pm$0.76 & - & - & 590$\pm$50 & 119$\pm$5 & G & Y \\
11071622-7723068 & 166.81758 & -77.38522 & - & - & - & - & - & G & Y \\
11071915-7603048 & 166.82979 & -76.05133 & 15.10$\pm$0.28 & 3607$\pm$114 & 0.884$\pm$0.008 & 595$\pm$32 & 371$\pm$8 & G & Y \\
11072022-7738111 & 166.83425 & -77.63642 & 14.91$\pm$0.50 & 3400$\pm$126 & 0.923$\pm$0.011 & 497$\pm$12 & 297$\pm$5 & G & Y \\
11072040-7729403 & 166.83500 & -77.49453 & 14.67$\pm$0.27 & 3296$\pm$64 & 0.921$\pm$0.004 & 594$\pm$4 & 115$\pm$5 & G & Y \\
11072825-7652119 & 166.86771 & -76.86997 & 15.12$\pm$0.27 & 3453$\pm$132 & 0.908$\pm$0.004 & 476$\pm$4 & 330$\pm$2 & G & Y \\
11073519-7734493 & 166.89663 & -77.58036 & 14.39$\pm$0.37 & 3397$\pm$20 & 0.896$\pm$0.012 & 620$\pm$17 & 110$\pm$4 & G & Y \\
11073832-7747168 & 166.90967 & -77.78800 & 13.82$\pm$0.42 & 3316$\pm$14 & 0.915$\pm$0.013 & 542$\pm$20 & 119$\pm$5 & G & Y \\
11074245-7733593 & 166.92687 & -77.56647 & 12.40$\pm$1.68 & - & - & 508$\pm$18 & 332$\pm$10 & G & Y \\
11074366-7739411 & 166.93192 & -77.66142 & 15.02$\pm$0.28 & 4140$\pm$157 & 0.969$\pm$0.005 & 362$\pm$14 & 351$\pm$7 & G & Y \\
11075225-7736569 & 166.96771 & -77.61581 & 11.98$\pm$1.29 & - & - & 564$\pm$65 & 128$\pm$5 & G & Y \\
11075588-7727257 & 166.98283 & -77.45714 & 17.36$\pm$0.40 & 4752$\pm$162 & - & 524$\pm$14 & 116$\pm$22 & U & Y \\
11075792-7738449 & 166.99133 & -77.64581 & 24.26$\pm$2.80 & - & - & 218$\pm$21 & 612$\pm$12 & G & Y \\
11075809-7742413 & 166.99204 & -77.71147 & 19.03$\pm$0.41 & 3423$\pm$116 & 0.901$\pm$0.012 & 191$\pm$34 & 501$\pm$10 & G & Y \\
11075993-7715317 & 166.99971 & -77.25881 & 11.66$\pm$3.07 & - & - & - & 290$\pm$17 & G & Y \\
11080002-7717304 & 167.00008 & -77.29178 & 15.62$\pm$0.70 & - & - & 574$\pm$18 & 112$\pm$4 & G & Y \\
11080148-7742288 & 167.00617 & -77.70800 & 10.65$\pm$0.40 & - & - & - & - & U & Y \\
11080297-7738425 & 167.01237 & -77.64514 & 17.69$\pm$0.50 & 4813$\pm$1391 & 0.955$\pm$0.009 & 316$\pm$9 & 432$\pm$9 & G & Y \\
11081509-7733531 & 167.06287 & -77.56475 & 14.13$\pm$0.28 & 4798$\pm$444 & 0.981$\pm$0.004 & 479$\pm$4 & 384$\pm$2 & G & Y \\
11081648-7744371 & 167.06867 & -77.74364 & 15.33$\pm$0.26 & 3409$\pm$72 & 0.914$\pm$0.005 & 587$\pm$4 & 114$\pm$3 & G & Y \\
11081703-7744118 & 167.07096 & -77.73661 & 14.64$\pm$0.58 & 3114$\pm$66 & 0.939$\pm$0.014 & 551$\pm$20 & 112$\pm$4 & G & Y \\
11082237-7730277 & 167.09321 & -77.50769 & 11.72$\pm$1.53 & - & - & 390$\pm$18 & 413$\pm$9 & G & Y \\
11082410-7741473 & 167.10042 & -77.69647 & 13.17$\pm$0.80 & - & - & 584$\pm$44 & 123$\pm$5 & G & Y \\
11083905-7716042 & 167.16271 & -77.26783 & 13.10$\pm$0.27 & 4264$\pm$303 & 0.948$\pm$0.003 & 564$\pm$3 & 474$\pm$10 & G & Y \\
11083952-7734166 & 167.16467 & -77.57128 & 13.18$\pm$0.84 & - & - & 417$\pm$16 & 318$\pm$8 & G & Y \\
11084069-7636078 & 167.16954 & -76.60217 & 13.06$\pm$0.26 & 3874$\pm$183 & 0.918$\pm$0.003 & 633$\pm$6 & 202$\pm$14 & G & Y \\
11084296-7743500 & 167.17900 & -77.73056 & - & - & - & - & - & G & Y \\
11085090-7625135 & 167.21208 & -76.42042 & 14.16$\pm$0.44 & 3251$\pm$75 & 0.945$\pm$0.011 & 413$\pm$23 & 308$\pm$7 & G & Y \\
11085242-7519027 & 167.21842 & -75.31742 & 14.09$\pm$0.28 & 3558$\pm$161 & 0.928$\pm$0.004 & 564$\pm$6 & 275$\pm$11 & G & N \\
11085367-7521359 & 167.22363 & -75.35997 & 14.79$\pm$0.27 & 3539$\pm$280 & 0.938$\pm$0.004 & 323$\pm$4 & 376$\pm$0 & G & Y \\
11085422-7732115 & 167.22592 & -77.53653 & 13.43$\pm$0.56 & - & - & 560$\pm$21 & 112$\pm$4 & G & Y \\
11085464-7702129 & 167.22767 & -77.03692 & 15.41$\pm$0.29 & 4025$\pm$275 & 0.916$\pm$0.005 & 409$\pm$47 & 466$\pm$20 & G & Y \\
11090512-7709580 & 167.27133 & -77.16611 & 13.08$\pm$4.23 & - & - & 647$\pm$44 & 238$\pm$82 & G & Y \\
11090915-7553477 & 167.28813 & -75.89658 & 14.09$\pm$0.32 & 3614$\pm$100 & 0.891$\pm$0.009 & 558$\pm$8 & 120$\pm$3 & G & N \\
11091172-7729124 & 167.29883 & -77.48678 & 14.76$\pm$0.40 & 3905$\pm$117 & - & 646$\pm$17 & 119$\pm$5 & U & Y \\
11091297-7729115 & 167.30404 & -77.48653 & 13.26$\pm$0.26 & 3763$\pm$110 & 0.894$\pm$0.004 & 610$\pm$4 & 131$\pm$4 & G & Y \\
11091380-7628396 & 167.30750 & -76.47767 & 14.92$\pm$0.95 & 3294$\pm$45 & - & 617$\pm$24 & 281$\pm$20 & G & Y \\
11091769-7627578 & 167.32371 & -76.46606 & 15.06$\pm$0.40 & 4541$\pm$171 & - & 559$\pm$10 & 133$\pm$4 & U & Y \\
11091812-7630292 & 167.32550 & -76.50811 & 16.15$\pm$0.46 & 3903$\pm$41 & 0.978$\pm$0.016 & 511$\pm$37 & 303$\pm$7 & G & Y \\
11092378-7623207 & 167.34908 & -76.38908 & - & 3990$\pm$123 & - & 65$\pm$8 & 566$\pm$20 & U & Y \\
11092855-7633281 & 167.36896 & -76.55781 & - & - & - & - & - & G & Y \\
11094006-7628392 & 167.41692 & -76.47756 & 15.20$\pm$0.26 & 3939$\pm$219 & 0.901$\pm$0.005 & 643$\pm$7 & 154$\pm$11 & G & Y \\
11094525-7740332 & 167.43854 & -77.67589 & 12.67$\pm$1.43 & - & - & 414$\pm$12 & 200$\pm$106 & G & Y \\
11094621-7634463 & 167.44254 & -76.57953 & 14.34$\pm$0.58 & 3431$\pm$35 & 0.909$\pm$0.015 & 327$\pm$14 & 375$\pm$9 & G & Y \\
11095003-7636476 & 167.45846 & -76.61322 & 12.79$\pm$0.83 & - & - & 14$\pm$5 & - & G & Y \\
11095340-7634255 & 167.47250 & -76.57375 & -536.10$\pm$0.73 & - & - & 269$\pm$8 & 584$\pm$12 & G & Y \\
11095407-7629253 & 167.47529 & -76.49036 & 15.16$\pm$0.70 & - & - & 419$\pm$19 & 321$\pm$14 & G & Y \\
11095873-7737088 & 167.49471 & -77.61911 & -635.00$\pm$2.41 & 3376$\pm$113 & - & 240$\pm$16 & 465$\pm$1 & G & Y \\
11100010-7634578 & 167.50042 & -76.58272 & 25.00$\pm$0.30 & 4979$\pm$191 & - & 236$\pm$8 & 554$\pm$11 & G & Y \\
11100369-7633291 & 167.51538 & -76.55808 & 15.48$\pm$0.42 & 3891$\pm$45 & 0.930$\pm$0.012 & 285$\pm$11 & 440$\pm$12 & G & Y \\
11100469-7635452 & 167.51954 & -76.59589 & 15.65$\pm$0.26 & 4088$\pm$227 & 0.946$\pm$0.003 & 480$\pm$9 & 437$\pm$9 & G & Y \\
11100704-7629377 & 167.52933 & -76.49381 & 14.04$\pm$0.40 & 4624$\pm$168 & - & 467$\pm$12 & 448$\pm$9 & U & Y \\
11101141-7635292 & 167.54754 & -76.59144 & 17.14$\pm$0.35 & 4470$\pm$174 & 0.975$\pm$0.006 & 551$\pm$6 & 438$\pm$13 & G & Y \\
11101153-7733522 & 167.54804 & -77.56450 & 13.17$\pm$0.41 & 3302$\pm$46 & 0.915$\pm$0.008 & 549$\pm$24 & 122$\pm$4 & G & Y \\
11102852-7716596 & 167.61883 & -77.28322 & 14.97$\pm$0.28 & 3287$\pm$71 & 0.933$\pm$0.007 & 600$\pm$20 & 107$\pm$13 & G & Y \\
11103481-7722053 & 167.64504 & -77.36814 & 14.76$\pm$1.87 & - & - & 572$\pm$22 & - & G & Y \\
11103801-7732399 & 167.65838 & -77.54442 & 13.70$\pm$0.36 & 4842$\pm$377 & - & 399$\pm$4 & 701$\pm$14 & G & Y \\
11104959-7717517 & 167.70663 & -77.29769 & -435.00$\pm$2.29 & 3327$\pm$198 & - & 252$\pm$19 & 498$\pm$10 & G & Y \\
11105076-7718032 & 167.71150 & -77.30089 & 14.83$\pm$0.38 & 3339$\pm$16 & 0.884$\pm$0.013 & 617$\pm$19 & 119$\pm$5 & G & Y \\
11105333-7634319 & 167.72221 & -76.57553 & 49.84$\pm$0.97 & 3287$\pm$171 & - & 67$\pm$4 & 425$\pm$9 & G & Y \\
11105597-7645325 & 167.73321 & -76.75903 & 18.02$\pm$1.12 & - & - & 321$\pm$10 & 257$\pm$8 & G & Y \\
11112260-7705538 & 167.84417 & -77.09828 & 13.75$\pm$0.31 & 3342$\pm$66 & 0.920$\pm$0.008 & 584$\pm$7 & 124$\pm$11 & G & Y \\
11113474-7636211 & 167.89475 & -76.60586 & 14.91$\pm$0.25 & 3943$\pm$258 & 0.905$\pm$0.002 & 663$\pm$2 & 150$\pm$18 & G & Y \\
11113965-7620152 & 167.91521 & -76.33756 & 16.17$\pm$0.25 & 3468$\pm$205 & 0.949$\pm$0.002 & 223$\pm$33 & 309$\pm$7 & G & Y \\
11114632-7620092 & 167.94300 & -76.33589 & 16.38$\pm$0.40 & 4617$\pm$207 & - & 461$\pm$10 & 373$\pm$8 & U & Y \\
11115400-7619311 & 167.97500 & -76.32531 & 15.50$\pm$0.25 & 3738$\pm$80 & 0.901$\pm$0.002 & 653$\pm$23 & 225$\pm$4 & G & Y \\
11120327-7637034 & 168.01362 & -76.61761 & 13.94$\pm$0.85 & 3247$\pm$15 & - & 556$\pm$109 & 164$\pm$7 & G & Y \\
11120984-7634366 & 168.04100 & -76.57683 & 15.20$\pm$0.30 & 3275$\pm$51 & 0.962$\pm$0.005 & 406$\pm$5 & 338$\pm$7 & G & Y \\
11122441-7637064 & 168.10171 & -76.61844 & 15.16$\pm$0.29 & 4899$\pm$326 & - & 414$\pm$5 & 389$\pm$8 & G & Y \\
11124210-7658400 & 168.17542 & -76.97778 & 15.04$\pm$0.26 & 4365$\pm$235 & 0.955$\pm$0.003 & 501$\pm$24 & 180$\pm$7 & G & Y \\
11124268-7722230 & 168.17783 & -77.37306 & 14.18$\pm$0.40 & 5196$\pm$81 & - & 380$\pm$7 & - & U & Y \\
11124299-7637049 & 168.17912 & -76.61803 & 14.46$\pm$0.40 & 4706$\pm$221 & - & 475$\pm$9 & 107$\pm$6 & U & Y \\
11130450-7534369 & 168.26875 & -75.57692 & 15.15$\pm$0.50 & 3321$\pm$50 & 0.929$\pm$0.011 & 583$\pm$31 & 291$\pm$9 & G & N \\
11132446-7629227 & 168.35192 & -76.48964 & 15.25$\pm$0.27 & 3374$\pm$90 & 0.877$\pm$0.006 & 622$\pm$9 & 233$\pm$10 & G & Y \\
11132737-7634165 & 168.36404 & -76.57125 & 13.87$\pm$0.25 & 3775$\pm$106 & 0.874$\pm$0.002 & 559$\pm$11 & 151$\pm$16 & G & Y \\
11132971-7629012 & 168.37379 & -76.48367 & 16.40$\pm$0.27 & 3339$\pm$67 & 0.913$\pm$0.006 & 633$\pm$6 & 106$\pm$3 & G & Y \\
11133356-7635374 & 168.38983 & -76.59372 & 16.91$\pm$0.63 & 3299$\pm$47 & - & 522$\pm$67 & 254$\pm$14 & G & Y \\
11141565-7627364 & 168.56521 & -76.46011 & 16.92$\pm$0.42 & 3384$\pm$124 & - & 598$\pm$7 & 248$\pm$5 & G & Y \\
11142454-7733062 & 168.60225 & -77.55172 & 16.01$\pm$1.08 & - & - & 342$\pm$29 & 396$\pm$15 & G & Y \\
11145031-7733390 & 168.70962 & -77.56083 & 14.44$\pm$0.25 & 3729$\pm$100 & 0.889$\pm$0.003 & 636$\pm$3 & 137$\pm$4 & G & Y \\
11182024-7621576 & 169.58433 & -76.36600 & 13.82$\pm$0.40 & 4317$\pm$127 & - & 539$\pm$18 & 116$\pm$12 & U & Y \\
11213079-7633351 & 170.37829 & -76.55975 & 14.99$\pm$0.30 & 3632$\pm$102 & 0.910$\pm$0.005 & 628$\pm$5 & 147$\pm$16 & G & N \\
11242981-7554237 & 171.12421 & -75.90658 & 11.56$\pm$0.50 & 3300$\pm$46 & - & 495$\pm$10 & 163$\pm$5 & G & Y \\
11291261-7546263 & 172.30254 & -75.77397 & 15.20$\pm$0.40 & 4818$\pm$96 & - & 477$\pm$12 & 130$\pm$5 & U & Y \\
\hline
\end{longtable}
\tablefoottext{a}{For spectra with a signal-to-noise ratio lower than three we did not report any velocity.} \\ 
\tablefoottext{b}{Empirical gravity indicator defined by \cite{Damiani:2014}.}\\ 
\tablefoottext{c}{Width at 10\% of the peak of the H$\alpha$ line.}\\ 
\tablefoottext{d}{The letters "G" and "U" indicate GIRAFFE and UVES, respectively.}\\
\tablefoottext{e}{The letters "Y" and "N" indicate the star is a known member or not, respectively}\\
\tablefoottext{f}{This star is likely a member of the $\epsilon$ Cha association.}
}

%% file: Tab2_elec_bis.tex
\onllongtab{
\begin{longtable}{cccccc}
\caption{Known cluster members from the literature and new member used to study the structural properties of the cluster.}\\
\hline \hline
RA & DEC & $T_{eff}$ & $Log(L_{Bol}/L_{\sun}$) & prov\tablefootmark{a} & Mass \\
(J2000) & (J2000) & (K) &  &  & $M_{\sun}$ \\
\hline
\endfirsthead
\caption{continued.}\\
\hline \hline
RA & DEC & $T_{eff}$ & $Log(L_{Bol}/L_{\sun}$) & prov\tablefootmark{a} & Mass \\
(J2000) & (J2000) & (K) &  &  & $M_{\sun}$ \\
\hline
\endhead
\hline
\endfoot
161.65812 & -77.60097 & 7200 & 0.95 & L & 1.64 \\
163.15392 & -74.67464 & 3161 & -1.00 & L & 0.14 \\
163.41575 & -77.20939 & 3451 & -1.49 & L & 0.30 \\
163.79017 & -77.51500 & 3198 & -1.08 & L & 0.15 \\
163.99887 & -77.41108 & 3640 & -0.36 & G & 0.37 \\
163.99887 & -77.41108 & 3640 & -0.36 & G & 0.37 \\
164.06825 & -76.51472 & 3044 & -1.51 & L & 0.09 \\
164.12683 & -77.19425 & 4350 & 0.06 & G & 0.85 \\
164.42579 & -76.99322 & 3451 & -0.33 & G & 0.28 \\
164.47400 & -77.41375 & 3426 & -0.94 & G & 0.28 \\
164.52487 & -77.19725 & 3091 & -1.66 & L & 0.10 \\
164.56987 & -77.28806 & 5250 & 0.28 & L & 1.43 \\
164.75450 & -77.37797 & 4135 & -0.04 & G & 0.68 \\
164.77912 & -77.02789 & 4980 & 0.66 & G & 1.71 \\
165.16758 & -76.32444 & 3306 & -1.11 & L & 0.20 \\
165.30708 & -77.37742 & 3091 & -1.60 & L & 0.10 \\
165.32812 & -76.45069 & 3940 & -0.24 & G & 0.56 \\
165.58029 & -75.61600 & 3198 & -1.21 & L & 0.15 \\
165.60379 & -77.55992 & 4519 & 0.22 & G & 1.01 \\
165.60875 & -75.04464 & 3161 & -1.12 & L & 0.14 \\
165.63604 & -77.48692 & 3610 & -0.73 & G & 0.40 \\
165.67429 & -77.40681 & 3125 & -1.31 & L & 0.12 \\
165.92442 & -77.44778 & 3058 & -1.44 & L & 0.10 \\
165.94850 & -77.33231 & 3125 & -0.74 & L & 0.13 \\
165.98675 & -77.35914 & 3442 & -0.56 & G & 0.29 \\
166.01771 & -76.65911 & 3234 & -1.35 & L & 0.16 \\
166.03787 & -76.45536 & 4350 & -0.02 & L & 0.88 \\
166.09479 & -77.30222 & 4060 & -1.89 & L & - \\
166.17742 & -77.69919 & 3270 & -1.03 & L & 0.19 \\
166.21250 & -76.42333 & 4571 & -0.26 & G & 1.01 \\
166.22021 & -76.43094 & 4031 & -0.51 & G & 0.72 \\
166.28133 & -78.20175 & 3161 & -1.11 & L & 0.14 \\
166.31113 & -77.19139 & 3433 & -0.49 & G & 0.28 \\
166.34467 & -77.15806 & 3161 & -1.46 & L & 0.13 \\
166.35300 & -76.43914 & 3576 & -0.90 & G & 0.38 \\
166.42917 & -77.44769 & 3223 & -0.93 & G & 0.17 \\
166.46921 & -76.30708 & 4405 & -0.23 & G & 0.96 \\
166.49083 & -76.13022 & 13500 & 1.99 & L & 3.00 \\
166.50042 & -75.12367 & 3198 & -1.33 & L & 0.14 \\
166.56417 & -77.36575 & 5770 & 1.20 & L & 2.53 \\
166.60642 & -76.56161 & 3091 & -1.28 & L & 0.11 \\
166.66437 & -77.60144 & 3091 & -2.13 & L & 0.10 \\
166.67417 & -76.59692 & 3273 & -0.81 & G & 0.19 \\
166.68108 & -77.44286 & 3415 & -0.38 & L & 0.26 \\
166.68792 & -77.45064 & 4343 & 0.12 & G & 0.83 \\
166.73888 & -77.70294 & 3435 & -0.65 & G & 0.29 \\
166.74608 & -77.31486 & 3234 & -0.96 & L & 0.17 \\
166.74746 & -75.51553 & 3091 & -1.92 & L & 0.10 \\
166.76667 & -76.52917 & 3198 & -1.24 & L & 0.14 \\
166.79783 & -77.77761 & 3708 & -0.40 & G & 0.42 \\
166.79921 & -76.43058 & 3091 & -1.74 & L & 0.10 \\
166.80025 & -76.53978 & 3979 & -0.31 & G & 0.62 \\
166.82979 & -76.05133 & 3606 & -0.53 & G & 0.38 \\
166.83425 & -77.63642 & 3399 & 0.28 & G & - \\
166.83500 & -77.49453 & 3295 & -0.88 & G & 0.20 \\
166.83642 & -77.63536 & 5860 & 1.08 & L & 2.20 \\
166.85179 & -77.73025 & 3024 & -1.44 & L & 0.09 \\
166.86771 & -76.86997 & 3452 & -0.56 & G & 0.29 \\
166.89663 & -77.58036 & 3397 & -1.03 & G & 0.26 \\
166.90967 & -77.78800 & 3316 & -1.08 & G & 0.21 \\
166.91000 & -75.88108 & 3161 & -1.47 & L & 0.13 \\
166.92687 & -77.56647 & 3091 & -0.96 & L & 0.11 \\
166.94208 & -77.66914 & 3024 & -1.43 & L & 0.09 \\
166.94400 & -76.25483 & 3024 & -1.89 & L & 0.08 \\
166.96771 & -77.61581 & 3058 & -1.15 & L & 0.10 \\
166.99971 & -77.25881 & 3024 & -1.24 & L & 0.09 \\
167.00617 & -77.70800 & 3955 & 0.48 & L & - \\
167.01371 & -77.65483 & 10010 & 1.85 & L & 2.70 \\
167.06287 & -77.56475 & 4798 & 0.83 & G & - \\
167.06867 & -77.74364 & 3409 & -0.65 & G & 0.27 \\
167.07096 & -77.73661 & 3113 & -1.02 & G & 0.12 \\
167.07900 & -77.65472 & 3058 & -1.26 & L & 0.10 \\
167.08075 & -77.53117 & 3161 & -3.21 & L & - \\
167.09325 & -77.50769 & 3125 & -1.06 & L & 0.12 \\
167.10042 & -77.69647 & 3058 & -1.09 & L & 0.10 \\
167.11042 & -77.26528 & 3024 & -1.72 & L & 0.09 \\
167.16467 & -77.57128 & 3024 & -1.24 & L & 0.09 \\
167.21208 & -76.42042 & 3250 & -1.30 & G & 0.17 \\
167.21842 & -75.31742 & 3558 & -0.41 & G & 0.34 \\
167.22363 & -75.35997 & 3538 & -0.50 & G & 0.33 \\
167.22587 & -77.53653 & 3091 & -1.18 & L & 0.11 \\
167.22767 & -77.03692 & 4024 & -0.35 & G & 0.67 \\
167.22904 & -76.54472 & 3058 & -1.46 & L & 0.10 \\
167.27133 & -77.16611 & 3161 & -1.17 & L & 0.13 \\
167.28813 & -75.89658 & 3614 & -0.90 & G & 0.41 \\
167.29883 & -77.48678 & 3905 & -0.17 & G & 0.52 \\
167.30125 & -77.48667 & 3560 & -0.48 & L & 0.34 \\
167.30125 & -77.48667 & 3415 & -0.58 & L & 0.27 \\
167.30404 & -77.48653 & 3762 & -0.69 & G & 0.51 \\
167.30750 & -76.47767 & 3294 & -1.16 & G & 0.19 \\
167.32371 & -76.46606 & 4541 & 0.24 & G & 1.03 \\
167.34908 & -76.38908 & 3990 & 0.19 & G & - \\
167.37138 & -76.98833 & 3091 & -1.62 & L & 0.10 \\
167.41692 & -76.47756 & 3939 & -0.15 & G & 0.54 \\
167.43854 & -77.67589 & 3024 & -1.24 & L & 0.09 \\
167.44254 & -76.57953 & 3430 & -0.52 & G & 0.28 \\
167.45492 & -77.52214 & 3058 & -1.44 & L & 0.10 \\
167.45846 & -76.61322 & 10500 & 1.76 & L & 2.80 \\
167.46925 & -77.67633 & 3415 & -2.54 & L & - \\
167.47887 & -76.58614 & 3024 & -2.85 & L & - \\
167.49471 & -77.61911 & 3376 & 0.15 & G & - \\
167.51954 & -76.59589 & 4087 & -0.20 & G & 0.68 \\
167.52933 & -76.49381 & 4623 & 0.30 & G & 1.12 \\
167.54804 & -77.56450 & 3301 & -0.88 & G & 0.21 \\
167.61883 & -77.28322 & 3286 & -1.12 & G & 0.19 \\
167.70663 & -77.29769 & 3327 & -0.22 & G & - \\
167.71150 & -77.30089 & 3338 & -0.88 & G & 0.23 \\
167.72221 & -76.57553 & 3287 & -0.72 & G & 0.20 \\
167.73321 & -76.75903 & 3024 & -0.89 & L & - \\
167.79513 & -76.69928 & 3488 & -2.52 & L & - \\
167.84417 & -77.09828 & 3341 & -0.98 & G & 0.23 \\
167.89475 & -76.60586 & 3942 & -0.54 & G & 0.65 \\
167.91521 & -76.33756 & 3467 & -0.23 & G & 0.28 \\
167.94300 & -76.33589 & 4617 & 0.47 & G & 1.10 \\
167.97500 & -76.32531 & 3738 & -0.44 & G & 0.45 \\
168.01362 & -76.61761 & 3246 & -1.11 & G & 0.17 \\
168.01462 & -77.43358 & 3161 & -1.52 & L & 0.12 \\
168.04100 & -76.57683 & 3275 & -0.64 & G & 0.20 \\
168.10171 & -76.61844 & 4898 & 0.42 & G & 1.49 \\
168.11550 & -76.73953 & 5410 & 0.70 & L & 1.97 \\
168.12883 & -76.74003 & 3705 & -0.41 & L & 0.42 \\
168.17542 & -76.97778 & 4364 & -0.33 & G & 0.93 \\
168.17783 & -77.37306 & 5196 & 0.62 & G & 1.88 \\
168.17912 & -76.61803 & 4706 & -0.08 & G & 1.13 \\
168.20254 & -76.78517 & 3270 & -1.28 & L & 0.18 \\
168.26875 & -75.57692 & 3321 & -1.10 & G & 0.21 \\
168.33383 & -77.01789 & 3451 & -0.74 & L & 0.30 \\
168.35192 & -76.48964 & 3373 & -1.05 & G & 0.25 \\
168.36404 & -76.57125 & 3775 & -0.52 & G & 0.49 \\
168.37379 & -76.48367 & 3338 & -0.97 & G & 0.23 \\
168.38983 & -76.59372 & 3299 & -1.08 & G & 0.20 \\
168.56521 & -76.46011 & 3383 & -0.73 & G & 0.26 \\
168.60225 & -77.55172 & 3270 & -0.96 & L & 0.19 \\
168.60879 & -77.55117 & 3024 & -1.38 & L & 0.09 \\
168.62108 & -76.42775 & 3161 & -1.43 & L & 0.13 \\
168.70962 & -77.56083 & 3728 & -0.49 & G & 0.45 \\
168.84083 & -77.40117 & 3161 & -1.02 & L & 0.14 \\
168.99279 & -77.48461 & 3198 & -1.24 & L & 0.14 \\
169.01196 & -76.41481 & 3955 & -2.47 & L & - \\
169.40417 & -77.07725 & 3778 & -0.41 & L & 0.47 \\
169.40800 & -76.77203 & 3024 & -1.89 & L & 0.08 \\
169.46712 & -76.49422 & 3198 & -1.06 & L & 0.15 \\
169.58154 & -76.36703 & 3524 & -0.82 & L & 0.34 \\
169.58433 & -76.36600 & 4317 & -0.11 & G & 0.88 \\
169.64079 & -76.71781 & 3125 & -1.01 & L & 0.12 \\
169.64883 & -79.59856 & 3161 & -0.68 & L & - \\
169.92558 & -76.39239 & 3125 & -1.48 & L & 0.11 \\
170.37829 & -76.55975 & 3632 & -0.65 & G & 0.40 \\
171.04942 & -76.51181 & 3125 & -1.33 & L & 0.12 \\
171.12421 & -75.90658 & 3299 & -0.70 & G & 0.21 \\
173.34696 & -76.36922 & 3198 & -0.70 & L & 0.17 \\
173.45525 & -76.31108 & 3198 & -1.24 & L & 0.14 \\
175.20696 & -74.99428 & 3024 & -1.57 & L & 0.09 \\
175.86121 & -78.07928 & 3125 & -1.02 & L & 0.12 \\
\hline
\end{longtable}
\tablefoot{\tablefoottext{a}{The letters G and L indicate that the data used for deriving the stellar mass 
are retrieved from the literature and the GES archive, respectively.}}
}